\documentclass[12pt]{article}
\usepackage{amssymb}
\usepackage{color}

\newtheorem{definition}{Definition}
\newtheorem{theorem}[definition]{Theorem}
\newtheorem{proposition}[definition]{Proposition}
\newtheorem{example}{Example}
\newtheorem{remark}{Remark}
\newtheorem{lemma}[definition]{Lemma}

\usepackage{bbm}

\font\rmsmall=cmr8

\newcommand\unit{\mathbbm{1}} 
\newcommand\re{\Re \kern-1.4truept e}
\newcommand\im{\Im \kern-1.4truept m}
\newcommand\Ran{{\rm{Ran\,}}}
\newcommand\Iso{{\rm{Iso}}}
\newcommand\M{\mathcal{M}}
\newcommand{\D}{\mathcal{D}} 
\newcommand{\h}{\mathsf{h}} 
\newcommand{\kk}{\mathsf{k}} 
\newcommand{\mm}{\mathsf{m}}
\newcommand{\T}{\mathcal{T}} 

\newcommand{\PP}{\mathbb{P}} 

\newcommand{\wT}{\widetilde{\T}}
\newcommand{\dT}{\T^{\mathsf{da}}}
\newcommand{\dL}{\Ll^{\mathsf{da}}}
\newcommand{\dfT}{\T^{\mathsf{df}}}
\newcommand{\dfL}{\Ll^{\mathsf{df}}}
\newcommand{\Ll}{\mathcal{L}} 
\newcommand{\Pmin}[1]{{\mathcal{P}_{\rm min}}(#1)}
\newcommand{\Z}[1]{{\mathcal{Z}}(#1)}  
\newcommand{\nT}{\mathcal{N(\mathcal{T})}} 
\newcommand{\FT}{\mathcal{F(\mathcal{T})}} 
\newcommand{\B}{\mathcal{B}(\h) } 
\newcommand{\E}{\mathcal{E}} 
\newcommand{\ee}[2]{|{e_{#1}}\rangle\langle{e_{#2}}|} 
\newcommand{\kb}[2]{|{#1}\rangle\langle{#2}|}

\newcommand{\tr}[1]{{\rm tr }\left(#1\right)}
\newcommand{\ptr}[2]{\hbox{\rm tr}_{#1}\left(#2\right)}
\newcommand{\mi}{{\mathrm{i}}}

\begin{document}

\title{Structure of Uniformly Continuous Quantum Markov Semigroups}

\author{\small Julien Deschamps,
Franco Fagnola, Emanuela Sasso and Veronica Umanit\`a }

%

\date{ }

\maketitle

\abstract{The structure of uniformly continuous quantum Markov semigroups with
atomic decoherence-free subalgebra is established providing a natural
decomposition of a Markovian  open quantum system into its noiseless
(decoherence-free) and irreducible (ergodic) components. This leads
to a new characterisation of the structure of invariant states and a new
method for finding decoherence-free subsystems and subspaces.
Examples are presented  to illustrate these results.}

\bigskip



\section{Introduction}
Quantum Markov Semigroups (QMS)  $\T=(\T_t)_{t\ge 0}$ on
the von Neumann algebra $\B$ of all bounded operators on
a complex separable Hilbert space $\h$ describe the evolution
of open quantum systems in quantum optics and quantum
information processing.

The structure of uniformly  continuous QMS and their generators has
been analyzed by several authors starting from the early works
by Davies\cite{Davies}, Spohn\cite{Spohn}, Lindblad\cite{Lindb},
Christensen and Evans\cite{ChrEv} (see, for instance, \cite{AlLe,
Partha, SG} and the references therein). In most of these
investigations, concern has been focused on the structure
of the generator and the relationships between its algebraic
properties and structural properties of the underlying
open quantum system.

In recent years, there has been a growing interest in the use
of QMSs to model open quantum systems having subsystems
which are not affected by decoherence (see Lidar and Whaley \cite{LW}, Knill and Laflamme\cite{KL}, Olkiewicz\cite{BO,O,O2},
Ticozzi and Viola\cite{TV}, see also \cite{CSU2,CSU3,generic} and
the references therein). In these applications the QMS (in
the Heisenberg picture) acts as a semigroup of automorphisms
of a von Neumann subalgebra $\nT$ of $\B$, called the
\emph{decoherence-free subalgebra}. This subalgebra
allows identification of noise protected subsystems where
states evolve unitarily, moreover, its structure and
relationship with the set of fixed points also has important
consequences on the asymptotic behaviour of the QMS
(see  \cite{DFR,FFRR,FV82,Spohn}).

In this paper, exploiting the explicit structure of purely atomic
von Neumann algebras, we give a full description of the structure
of  uniformly continuous QMSs with atomic decoherence-free subalgebra.

Our first result, Theorem \ref{th:NT-factor}, shows that, when $\nT$
is a type I factor, the Hilbert space $\h$ is (isomorphic to) the tensor
product of two Hilbert spaces  $\kk$ and $\mm$,
$\nT$ is isomorphic to $\mathcal{B}(\kk)\otimes
\unit_{\mm}$ where $\unit_{\mm}$
is the identity operator on $\mm$ and the operators
in a Gorini-Kossakowski-Sudarshan-Lindblad (GKSL) representation
of the generator factorise accordingly. Linear maps
$\T_t$ (up to unitary isomorphism) factorise as
$\T_t^\kk\otimes\T_t^\mm$ where
$\T^\kk$ and $\T^\mm$ are QMS on $\mathcal{B}(\kk)$
and  $\mathcal{B}(\mm)$ respectively and the QMS  $\T^\kk$
acts as a semigroup of automorphisms
($\T_t^\kk(x)=\hbox{\rm e}^{ \mi t K} x
\hbox{\rm e}^{ -\mi t K}$ for some self-adjoint $K$ on $\kk$).
In this way  the decoherence-free (noise protected) part of the
system turns out to be essentially independent of the noisy part
of the system. This result shows, roughly speaking,   that the
only way of maintaining a subsystem free from decoherence
is by keeping it isolated.

The main result, Theorem \ref{th:main}, concerns the case
where $\nT$ is an atomic algebra and so it is a direct sum  of
type I factors and the above considerations apply to each
term of the direct sum.

Our result has important consequences. The first concerns the
structure of all invariant states of QMSs with a faithful
invariant state, which is completely characterised by
Theorem \ref{th:form-eta}  extending to infinite dimensional
Hilbert space $\h$ a result by Baumgartner and Narnhofer  \cite{BN}.
The second,  Theorem \ref{th:EID}, is a simple sufficient condition for
establishing environment induced decoherence (\cite{O,O2, CSU2}).
Moreover, the decomposition
$\h=\oplus_{i\in I}\left(\kk_i\otimes\mm_i\right)$
of the Hilbert space $\h$ as in Theorem \ref{th:main} allows us
 to identify immediately  decoherence-free quantum subsystems,
in the sense of Ticozzi and Viola \cite{TV}, and decoherence-free subspaces, as defined by Lidar et al. \cite{LW}  (see also
\cite{AFR}), of a given quantum Markovian system.

The decoherence-free subalgebra plays a key role in all the
above decompositions. Indeed, starting from the leading idea
that an atomic subalgebra has a special structure, we undertake
the analysis of the structure of the generator of an arbitrary
uniformly  continuous QMS $\T$ on $\B$ and find an infinite
dimensional generalisation of all the main known results.

The structure of the paper is as follows.
In section \ref{sect:df-subalg} we recall the definitions and
review some basic properties of the decoherence-free
subalgebra $\nT$ and the set of fixed points $\FT$ for the maps
$\T_t$. In order to make our exposition self-contained we collect
there several preliminary results scattered in the literature.
Moreover, we also prove (Proposition \ref{prop:aut}) that
the center of $\nT$ is contained in $\FT$.
In section \ref{sect:struct-QMS} we establish  our main
results Theorems \ref{th:NT-factor} and \ref{th:main}.
In the next section we prove our result
on the structure of invariant states also deducing spectral
properties of the Hamiltonian $K$ in the decoherence-free
part of the QMS (Lemma \ref{lem:K-pp-spec}) and showing
that the decoherence-free subalgebra of irreducible QMSs
is trivial (Proposition \ref{prop:Tirr-NTtriv}).
In section \ref{sect:appl} we discuss the applications to
environment induced decoherence and decoherence-free
subsystems and subspaces.
In the final section we present two examples, generic and circulant
QMSs, to illustrate how our results work in a  concrete set-up.
The appendix discusses a known characterisation of atomic
von Neumann algebras that we have been unable to find in
the wealthy literature on the subject.

\section{The decoherence-free subalgebra  of a QMS}
\label{sect:df-subalg}

Let $\h$ be a complex separable Hilbert space. A QMS on
the algebra $\B$ of all bounded operators on $\h$ is a semigroup
$\T=(\T_t)_{t\ge 0}$ of completely positive, identity preserving
normal maps which is also weakly$^*$ continuous. We will make
the assumption from now on that $\T$ is indeed uniformly  continuous
i.e. $\lim_{t\to 0^+} \sup_{\Vert x\Vert \le 1}
\left\Vert \T_t(x)-x\right\Vert=0$. Its generator $\Ll$  can be
represented in the well-known  (see \cite{Partha,SG})
 Gorini-Kossakowski-Sudarshan-Lindblad (GKSL)  form as
\begin{equation}\label{eq:GKSL}
\Ll(x)=\mi [H,x]-\frac{1}{2}\sum_{\ell\geq 1}
\left(L_\ell^*L_\ell x-2L_\ell^*xL_\ell+xL_\ell^*L_\ell\right),
\end{equation}
where $H=H^*$ and $(L_\ell)_{\ell\geq 1}$ are operators on
$\h$ such that the series $\sum_{\ell\geq 1}L^*_\ell L_\ell$ is
strongly convergent  and $[\cdot,\cdot]$ denotes the
commutator $[x,y]=xy-yx$.
The choice of operators $H$ and
$(L_\ell)_{\ell\geq 1}$ is not unique (see Parthasarathy
\cite{Partha} Theorem 30.16), however, this will not create
any inconvenience in this paper.

Given a GKSL representation of $\Ll$ we call $\Ll_0$
\[
\Ll_0(x):=-\frac{1}{2}\sum_{\ell\geq 1}
\left(L_\ell^*L_\ell x-2L_\ell^*xL\ell+xL_\ell^*L_\ell\right),
\qquad x\in\B,
\]
dissipative part of $\Ll$ and $\mi \delta_H(x):=\mi[H,x]$
Hamiltonian part  of $\Ll$ by abuse of language. Clearly, we
have $\Ll=\mi\delta_H+\Ll_0$.

The \emph{decoherence-free (DF) subalgebra} of $\T$ is
defined by
\begin{equation}\label{eq:NT-def}
\nT=\{x\in\B\,\mid\,\T_t(x^*x)=\T_t(x)^*\T_t(x),\ \T_t(xx^*)=\T_t(x)\T_t(x)^*\ \ \forall\,t\geq 0\}.
\end{equation}
It is a well known fact that $\nT$ is the biggest von Neumann
subalgebra of $\B$ on which every $\T_t$ acts as a
$*$-automorphism by the following (see e.g.
Evans\cite{Evans} Theorem 3.1, \cite{DFR}  Proposition 2.1).

\begin{proposition}\label{prop-struct-NT}
Let $\T$ be a quantum Markov semigroup on $\B$ and let
${\mathcal{N}}(\T)$ be the set defined by (\ref{eq:NT-def}).
Then
\begin{enumerate}
\item $\nT$ is $\T_t$-invariant for all $t\ge 0$,
\item for all $x\in\nT$, $y\in\B$ and $t\geq 0$ we have
$\T_t(x^*y)=\T_t(x^*)\T_t(y)$ and
$\T_t(y^*x)=\T_t(y^*)\T_t(x)$,
\item $\nT$ is a von Neumann subalgebra of $\B$.
\end{enumerate}
\end{proposition}

\noindent{\bf Proof. }
(1) Let $x\in \nT$ and $t>0$.  For all $s>0$ we have
\[
\T_s\left(\T_t(x^*x)\right) = \T_{s+t}(x^*x)
= \T_{s+t}(x^*)\T_{s+t}(x)
= \T_s\left(\T_t(x)^*\right)\T_s\left(\T_t(x)\right).
\]
Exchanging $x$ and $x^*$ we find the identity
\[
\T_s\left(\T_t(xx^*)\right)
=\T_s\left(\T_t(x)\right)\T_s\left(\T_t(x)^*\right).
\]
Thus, $\T_t(x)$ belongs to $\nT$.

(2) For all $t\ge 0$ and $x,y\in\B$ define
${\mathcal{D}}_t(x,y)=\T_t(x^*y)-\T_t(x^*)\T_t(y)$.
For every state $\omega$ on $\B$ and every complex
number $z$, by the  complete positivity of $\T_t$, we have
$\omega\left({\mathcal{D}}_t(zx+y,zx+y)\right)\ge 0$.
Now, if $x\in\nT$, then
$\omega\left({\mathcal{D}}_t(x,x) \right)=0$ so that
\[
0 \le \omega\left({\mathcal{D}}_t(zx+y,zx+y)\right)
= 2\re\left( \bar{z}\omega\left({\mathcal{D}}_t(x,y)\right)\right)
+ \omega\left({\mathcal{ D}}_t(y,y)\right)
\]
for all complex number $z$. It follows that
$\omega\left({\mathcal{D}}_t(x,y)\right)=0$
i.e. $\T_t(x^*y)=\T_t(x^*)\T_t(y)$, by the arbitrarity of
$\omega$, and (2) is proved.

(3) $\nT$ is a vector space by (2).
Moreover, for all $x,y\in\nT$, we have
\[
\T_t((xy)^*(xy)) = \T_t(y^*)\T_t(x^*)\T_t(x)\T_t(y)
=\T_t((xy)^*)\T_t(xy).
\]
The invariance of ${\mathcal{N}}(\T)$ for the adjoint
is obvious.  Finally, for any net $(x_\gamma)_{\gamma}$
of elements of ${\mathcal{N}}(\T)$ converging $\sigma$-strongly
to a $x$ in $\B$ we have
\[
\T_t(x^*x)=\lim_\gamma \T_t(x^*x_\gamma)
=\lim_\gamma \T_t(x^*)\T_t(x_\gamma)=\T_t(x^*)\T_t(x).
\]
Therefore $x$ belongs to $\nT$ and (3) is proved.
\hfill $\square$ \medskip

The decoherence-free subalgebra $\nT$ can be characterised
as follows.

\begin{proposition}\label{prop:NT-aut}
For all  self-adjoint $H$ in a GKSL representation of the $\Ll$
as in (\ref{eq:GKSL}) we have
\[
\nT \subseteq \left\{x\in\B\,\mid\,\T_t(x)
=\hbox{\emph{e}}^{\mi tH}x\,\hbox{\emph{e}}^{-\mi tH}\ \
\forall\,t\geq 0\right\}.
\]
\end{proposition}

\noindent{\bf Proof.} 
If $x$ belongs to $\nT$, then, differentiating the identity
$\T_t(x^*x)=\T_t(x^*)\T_t(x)$ at $t=0$, we have $\Ll(x^*x)=x^*\Ll(x)+\Ll(x^*)x$.
Computing
\begin{equation}\label{eq:Lxstarx}
\Ll(x^*x) -x^*\Ll(x) - \Ll(x^*) x =
\sum_{\ell\ge 1} \left[L_\ell, x\right]^*\left[L_\ell, x\right]
\end{equation}
for an arbitrary $x\in\B$, we find $\left[L_\ell, x\right]=0$
for $x\in \nT$. Moreover, since $\nT$ is a $^*$-algebra,  $x^*\in\nT$
so that $\left[L_\ell, x^*\right]=0$ and, taking the adjoint,
$\left[L_\ell^*, x\right]=0$. It follows that
$\Ll(x)=\mi \left[H, x\right]$ for all $x\in \nT$.

Now fix $t>0$ and $x\in\nT$. For all $0\le s \le t$, $\T_s(x) \in\nT$ and,
differentiating with respect to $s$,
\begin{eqnarray*}
\frac{d}{ds}\,\hbox{\rm e}^{\mi (t-s)H}\T_s(x)
\hbox{\rm e}^{-\mi (t-s)H}
& = & \mi  \hbox{\rm e}^{\mi (t-s)H} \left[H, \T_s(x)\right]
\hbox{\rm e}^{-\mi (t-s)H} \\
& - &\mi \hbox{\rm e}^{\mi (t-s)H} H \T_s(x)
\hbox{\rm e}^{-\mi (t-s)H}  \\
& - & \mi \hbox{\rm e}^{\mi (t-s)H} \T_s(x) H
\hbox{\rm e}^{-\mi (t-s)H} \\
& = & 0.
\end{eqnarray*}
We thus deduce that the function  $s\mapsto
\hbox{\rm e}^{\mi (t-s)H}\T_s(x)
\hbox{\rm e}^{-\mi (t-s)H}$ is
constant on $[0,t]$, and taking its values at $s=0$ and $s=t$,
we find  $\T_t(x)=\hbox{\rm e}^{\mi tH}x
\hbox{\rm e}^{-\mi tH} $.
\hfill $\square$ \medskip

In addition, we can characterise (\cite{FFRR} Theorem 2.1) $\nT$
in terms of operators $H,L_\ell$ in any GKSL representation of $\Ll$.
First define iterated commutators $\delta^{n}_H(X)$
recursively by $\delta^{0}_H(X)=X, \delta^{1}_H(X)=[H,X]$,
$\delta^{n+1}_H(X) = [H, \delta^{n}_H(X)]$.

\begin{proposition}\label{prop:NT-comm}
The decoherence-free subalgebra $\nT$ is the commutant of the set
of operators
\begin{equation}\label{eq:iter-comm}
\left\{\delta_H^{n}(L_\ell),\delta_H^{n}(L_\ell^*)\,
\mid\,n\geq 0, \ell\geq 1\right\}.
\end{equation}
\end{proposition}

\noindent{\bf Proof.}
If $x\in\nT$, then $\T_t(x)\in \nT$ by Proposition \ref{prop-struct-NT}
(1), and so $\Ll(x)=\lim_{t\to 0^+}t^{-1}(\T_t(x)-x)\in\nT$. Moreover,
arguing as in the proof of Proposition \ref{prop:NT-aut}, we find
$[L_\ell,x]=0=[L_\ell^*,x]$ and $\Ll(x) = \mi [H,x]
=\mi\delta_H(x)\in\nT$.
We now proceed by induction. Clearly all elements of $\nT$
commute with $\delta^{0}_H(L_\ell)=L_\ell$ and $\delta^{0}_H(L_\ell^*)=L_\ell^*$.
Suppose that they commute with $\delta^n_H(L_\ell)$
and $\delta^n_H(L_\ell^*)$ for some $n$, then, by the
Jacobi identity
\[
\left[  x, \delta^{n+1}_H(L_\ell)\right]
= - \left[  H, \left[ \delta^{n}_H(L_\ell),x\right]\right]
- \left[  \delta^{n}_H(L_\ell), [x,H]\right] =0
\]
because $[x,H]=\mi\Ll(x)\in\nT$. Thus, all elements of $\nT$
commute with $\delta^{n+1}_H(L_\ell)$ and, also with
$\delta_H^{n+1}(L_\ell^*)=-\delta_H^{n+1}(L_\ell)^*$
because $\nT$ is a $^*$-algebra. This shows that $\nT$
is contained in the commutant of the set (\ref{eq:iter-comm}).

Conversely, if $x$ belongs to the commutant of the set
(\ref{eq:iter-comm}), then it commutes with $L_\ell, L_\ell^*$
and so $\Ll(x) = \mi \delta_H(x)$. Moreover, $\delta_H(x)$
commutes with $L_\ell$ and $L_\ell^*$ because, by the
Jacobi identity, $[L_\ell,\delta_H(x)] = - [H,[x,L_\ell]]
-[x,\delta_H(L_\ell)]=0$ and, similarly,  $[L_\ell^*,\delta_H(x)] =0$.
Suppose, by induction, that $\Ll^n(x)=\mi^n\delta_H^n(x)$
and $\delta_H^k(x)$ commutes with $\delta_H^{n-k}(L_\ell)$
and $\delta_H^{n-k}(L_\ell^*)$ for all $k\le n$, for some $n$.
Then, $\Ll^{n+1}(x)=\mi^n\Ll(\delta_H^n(x))$ and so
\begin{eqnarray*}
\Ll^{n+1}(x) & = & \mi^{n+1}\delta_H^{n+1}(x)
+\frac{1}{2}\sum_{\ell\ge 1}\left(
L_\ell^*\left[\delta_H^n(x), L_\ell\right]
+\left[L_\ell^*,\delta_H^n(x)\right] L_\ell\right)  \\
& = & \mi^{n+1}\delta_H^{n+1}(x)
\end{eqnarray*}
Moreover, by repeated use of the Jacobi identity, we have
\begin{eqnarray*}
\left[ \delta_H^k(x), \delta_H^{n+1-k}(L_\ell)  \right]
& = & -\left[ \left[ \delta_H^{k-1}(x),
\delta_H^{n+1-k}(L_\ell)\right],H\right]    \\
& - & \left[ \left[ \delta_H^{n+1-k}(L_\ell),H\right],
\delta_H^{k-1}(x)\right] \\
& = & \left[ \delta_H^{k-1}(x),\delta_H^{n+2-k}(L_\ell)\right] \\
& = & \dots = \left[ x,  \delta_H^{n+1}(L_\ell)\right] = 0,
\end{eqnarray*}
and, similarly, $\left[ \delta_H^k(x), \delta_H^{n+1-k}(L_\ell^*)  \right]=0$.
It follows that $\Ll^n(x) = \mi^n\delta_H^n(x)$ for all $n\ge 0$
and so $\T_t(x) = \hbox{\emph{e}}^{\mi tH}x\,\hbox{\emph{e}}^{-\mi tH}$,  thus $x\in\nT$ by
Proposition \ref{prop:NT-aut}.
\hfill $\square$ \medskip

It is worth noticing here that Proposition \ref{prop:NT-comm}
holds for any GKSL representation of the generator $\Ll$.
Indeed, even if the operators $L_\ell, H$ are not uniquely
determined by $\Ll$ (see \cite{Partha} Theorem 30.6)
all other possible choices are of the form
\[
L_\ell^\prime = \sum_{m} u_{\ell m}L_m + z_m \unit, \qquad
H^\prime = H + c + \frac{1}{2\mi}\left( X - X^*\right)
\]
where $(u_{\ell m})_{\ell,m\ge 1}$ is a unitary matrix,
$(z_m)_{m\ge 1}$ is a sequence of complex scalars such that
$\sum_{m}|z_m|^2 <\infty$, $c\in\mathbb{R}$ and
$X= \sum_{m,j} \overline{z}_m u_{m j} L_j$. As a
consequence, the commutant of the set of operators in
Proposition \ref{prop:NT-comm} does not change replacing
$L_\ell, H$ by $L_\ell^\prime, H$.

Propositions \ref{prop:NT-aut} and \ref{prop:NT-comm}
have been extended to weakly$^*$ continuous QMS with
generators in a generalised GKSL form in \cite{DFR}.

\medskip

Our investigation is concerned with the implications of the structure
of $\nT$, as a von Neumann subalgebra of $\B$, on the structure
of $\T$. Let $\unit_\kk$ denote the identity operator on a Hilbert
space $\kk$. We begin by recalling some basic definitions on
operator algebras (see Takesaki \cite{Take}).

\begin{definition}\label{def:op-alg}
Let $\M$ be a von Neumann algebra acting on $\h$.
\begin{itemize}
\item[(a)] The center $\Z{\M}$ of $\M$ is the von Neumann
algebra of  elements $x$ of $\M$ commuting with all $y\in\M$,
\item[(b)] $\M$ is a \emph{factor} if $\Z{\M}=\mathbb{C}\unit_\h$.
\item[(c)] $\M$ is a \emph{type I factor} if it is a factor and
possesses a non-zero minimal projection.
\end{itemize}
\end{definition}

Throughout the paper we  will assume that
\begin{equation}\label{eq:NT-decomp}
\nT = \oplus_{i\in I} p_i \nT p_i
\end{equation}
where $(p_i)_{i\in I}$ is an (at most countable) family of mutually
orthogonal non-zero projections, which are minimal projections
in the center of $\nT$, such that $\sum_{i\in I} p_i =\unit$ and
each von Neumann algebra $ p_i \nT p_i  $ is a type I factor.

It is known that this property characterises atomic von Neumann
algebras.  We include a proof in the Appendix for completeness.

\begin{proposition}\label{prop:aut}
Let $\M$ be an atomic von Neumann algebra and let
$(\alpha_t)_{t\ge 0}$ be a weak* continuous semigroup
of $*$-automorphisms on $\M$. Then $\alpha_t(x)=x$ for all
$x\in\Z{\M}$ and $t\geq 0$.
\end{proposition}
\noindent{\bf Proof.}
Let  $(p_i)_{i\in I}$ be a family of mutually orthogonal projections
which are minimal in $\Z{\M}$ such that $\sum_{i\in I}p_i=\unit$.
Given $x\in\Z{\M}$, every $p_ixp_i$
belongs to $p_i\Z{\M}p_i=\mathbb{C}p_i$ by minimality; hence,
it is enough to prove that every $p_i$ is a fixed point for $\alpha$.

Since $\alpha_t$ is a $*$-automorphism, clearly
$\{\,\alpha_t(p_i)\,\mid\, i\in I\,\}$ is a family of mutually
orthogonal projections; in particular this family is contained
in $\Z{\M}$, because for all $x\in\M$ we have $x=\alpha_t(y)$
for some $y\in\M$, and so
\[
x\alpha_t(p_i)=\alpha_t(yp_i)=\alpha_t(p_iy)=\alpha_t(p_i)x
\]
for all $i\in I$. Moreover, every $p_j\alpha_t(p_i)p_j$ is clearly
a projection in $\Z{\M}$ for each $j\in I$, because
\[
(p_j\alpha_t(p_i)p_j)^2 = p_j\alpha_t(p_i)p_j \alpha_t(p_i) p_j
= p_j\alpha_t(p_i)^2 p_j=p_j \alpha_t(p_i) p_j
\]
and $(p_j\alpha_t(p_i)p_j)^*=p_j\alpha_t(p_i)p_j$. In addition,
we have also $p_j\alpha_t(p_i)p_j\le p_j$ since $\alpha_t(p_i) \le
\alpha_t(\unit)=\unit$. Therefore, by the minimality of projections $p_j$, for every $t\ge 0$, either $p_j\alpha_t(p_i)p_j=0$ or
$p_j\alpha_t(p_i)p_j=p_j$. By the weak$^*$ continuity
of the map $t\mapsto\alpha_t(p_i)$, we find
$p_i\alpha_t(p_i)p_i = p_i$ and $p_j\alpha_t(p_i)p_j=0$
for $j\not=i$. It follows that $p_i\alpha_t(p_i^\perp)p_i
=p_i\alpha_t(\unit-p_i )p_i =0$, so that $p_i\alpha_t(p_i^\perp)
=\alpha_t(p_i^\perp)p_i=0$, by the positivity of $\alpha_t(p_i)$,
and $\alpha_t(p_i)=p_i$.
\hfill $\square$ \medskip

We now study the structure of $\nT$.

\begin{proposition}\label{pi-inv}
If $\nT$ is an atomic algebra, then its center $ \Z{\nT}$ is
contained in the set of fixed points of all the maps $\T_t$.
\end{proposition}

\noindent{\bf Proof.}
We know from Proposition \ref{prop:NT-aut} that every $\T_t$
acts as a $*$-automorphism on $\nT$  with inverse  $x\mapsto
\hbox{\rm e}^{-\mi tH}x\,\hbox{\rm e}^{\mi tH}$.
Defining $\alpha_t$ as the restriction of $\T_t$ to $\nT$
for all $t\geq 0$, we obtain a weak* continuous group
of $*$-automorphism on $\nT$ and the conclusion follows
from Proposition \ref{prop:aut}.
\hfill $\square$ \medskip

 We now briefly recall some results on the set
\begin{equation}\label{eq:FT}
\FT=\left\{\, x\in\B \,\mid\, \T_t(x)=x, \, \forall t\ge 0  \right\}.
\end{equation}
of fixed points for the linear maps $\T_t$ that will be useful in the
sequel. Clearly, $\FT$ is a vector space containing $\unit$, and
$a\in\FT$ if and only if $a^*\in \FT$; moreover it
is norm-closed and weakly$^*$-closed, and so it is an
operator system.

\begin{lemma}\label{lem:proj-FT}
 An orthogonal projection $p\in \B$ belongs to $\FT$ if and only if
it commutes with the operators $L_\ell$ and $H$ of any GKSL representation of $\Ll$.
\end{lemma}

\noindent{\bf Proof.}
Clearly, if $p$ commutes with the operators $L_\ell$ and $H$,
then $\Ll(p) = 0$ and $\T_t(p) = p$ for all $t \ge 0$.

Conversely, if $\T_t(p)=p$ for all $t\ge 0$, then $\Ll(p)=0$.
Left and right multiplying by the orthogonal projection
$p^\perp=\unit-p$, we have
\[
0 = p^\perp \Ll(p) p^\perp =
p^\perp \sum_{\ell\ge 1}L^*_\ell p L_\ell p^\perp
\]
and so $p L_\ell p^\perp=0$. Similarly, starting from $\Ll(p^\perp)
= \Ll(\unit-p)  = \Ll(\unit)-\Ll(p)=0$, we find $p^\perp L_\ell p=0$.
Taking the adjoints we also obtain $p L_\ell^*p^\perp
=p^\perp L_\ell^* p=0$, and so $p$ commutes with $L_\ell$
and $L_\ell^*$. As a results $\Ll(p)=\mi [H,p]=0$ and
$p$ also commutes with $H$.
\hfill $\square$ \medskip

The following example shows that $\FT$, unlike $\nT$, may
not be an algebra.  We refer to \cite{CSU2}, section 4,
for additional examples.

\begin{example}\label{ex:FT-no-alg}{\rm
Let $\h=\mathbb{C}^3$ with canonical orthonormal basis
$(e_i)_{0\le i \le 2}$ and let $\Ll$ be the generator in the GKSL
form with a single non-zero operator $L=|e_0\rangle\langle e_2|$
and $H = L^*L = |e_2\rangle\langle e_2|$.
A straightforward computation yields, for a $3\times 3$
matrix $a=(a_{ij})_{0\le i,j\le 2}$ we have
\begin{eqnarray*}
\Ll(a) & = &  (a_{00}-a_{22}) |e_2\rangle\langle e_2|\\
 &- &\left(\frac{1}{2}+\mi\right)
\left(a_{02} |e_0\rangle\langle e_2|
+ a_{12} |e_1\rangle\langle e_2|\right) \\
& - & \left(\frac{1}{2}-\mi\right)
\left(a_{20} |e_2\rangle\langle e_0|
+ a_{21} |e_2\rangle\langle e_1|\right).
\end{eqnarray*}
Thus, $a$ is a fixed point for the QMS generated by $\Ll$ if and only
if $a_{02}=a_{12}=a_{20}=a_{21}=0$ and $ a_{00}=a_{22}$.
Now, it is easy to see that, for such an $a$, the matrix $a^*a$
also satisfies $\Ll(a^*a)=0$ if and only if, by (\ref{eq:Lxstarx}),
$a$ commutes with $L$, namely $a_{10}=0$.

Since by Proposition \ref{prop:NT-comm} every element in N(T) commutes with $L$ and $L^*$, another computation shows that,
if $a\in\nT$, then it commutes with $L$ and $L^*$, therefore $a_{ij}=0$ for $i\not=j$ and
$a_{00}=a_{22}$. In this case, since $\delta_H(L)=L$, it also
commutes also with all the iterated commutators
$\delta_H^n(L)=L,\delta_H^n(L^*)=L^*$.
In other words, $a$ belongs to $\nT$ if and only if $a_{ij}=0$ for $i\not=j$ and $a_{00}=a_{22}$. Hence, in this example  $\nT\subseteq\FT$.
}
\end{example}

In many situations, however, $\FT$ is an algebra and is
contained in $\nT$.  Further simple but useful properties (see
\cite{Evans, FFRR, FV82})  are collected in the following
proposition.

\begin{proposition}\label{prop:FT-alg}
The following hold:
\begin{enumerate}
\item the fixed points set $\FT$ is a $^*$-algebra if and only if
it is contained in the decoherence-free subalgebra $\nT$,
\item if the QMS $\T$ has a faithful normal invariant state,
then $\FT$ is a von Neumann subalgebra of $\B$,
\item if $\FT$ is a von Neumann subalgebra of $\B$, then
it coincides with the commutant of the set of operators
$\left\{\, L_\ell, L_\ell^*, H\,\mid \,\ell\ge 1\,\right\}$.
\end{enumerate}
\end{proposition}

\noindent{\bf Proof.} (1)  If $\FT$ is contained in $\nT$, then,
for all $x\in\FT\subseteq\nT$, we have $\T_t(x^*x)
=\T_t(x^*)\T_t(x)= x^*x$ and $x^*x\in \FT$.
Conversely, if $\FT$ is a $^*$-algebra, then, for
all $a\in\FT$, $a^*a\in \FT$, thus
$\T_t(a^*a)=a^* a = \T_t(a^*)\T_t(a)$ and
$a$ belongs to $\nT$.

(2)  Let $\rho$ be a faithful invariant state for $\T$.
If $\T_t(x)=x$ for all $t\ge 0$, then, by complete positivity
$x^*x=\T_t(x^*)\T_t(x)\le \T_t(x^* x)$, and
$\tr{\rho(\T_t(x^*x)-x^*x)}=0$ by the invariance of $\rho$.
Thus $\T_t(x^*x)=x^*x$ for all $t\ge 0$ because $\rho$ is
faithful and so $x^*x\in\FT$.

(3) If $\FT$ is a von Neumann subalgebra of $\B$, then it is
generated by its projections which belong to the commutant
of $\{L_\ell,L_\ell^*,H\,\mid\, \ell\ge 1\}$ by Lemma
\ref{lem:proj-FT}. Thus $\FT$ is contained in this commutant. Conversely, any $x$ commuting with $L_\ell,L_\ell^*,H$
satisfies $\Ll(x)=0$ and so $\T_t(x)=x$.
\hfill $\square$ \medskip


We finish this section by recalling two results on the asymptotic
behaviour of a QMS related with $\FT$ and $\nT$.
The first one follows from an application of the mean ergodic
theorem (see \cite{FV82} (Theorem 1.1)).

\begin{theorem}\label{th:mean-ergodic}
For a QMS  $\T$ with a faithful invariant state the limit
\[
\mathcal{E}(x) =  w^*-\lim_{t\to\infty}\frac{1}{t}
\int_0^t \T_s(x) ds
\]
exists for all $x\in\B$ and defines a $\T$-invariant normal
conditional expectation $\mathcal{E}$ onto the von Neumann
algebra $\mathcal{F}(\T)$ of fixed points for $\T$.
\end{theorem}

The second one, proved in \cite{FV82} Theorem 3.3,  ensures
that maps $\T_t$ converge to the above conditional
expectation as $t$ tends to infinity.

\begin{theorem}\label{th:FT=NT}
Suppose that there exists a faithful family of normal invariant
states for the QMS $\T$. Then $\FT=\nT$ implies that
\[
w^*-\lim_{t\to\infty} \T_t(x)  = \mathcal{E}(x)
\]
\end{theorem}

The idea behind this result is quite simple. If
$x\in\nT$, then $\T_t(x)
= \hbox{\rm e}^{\mi tH}x\,\hbox{\rm e}^{-\mi tH}$,
thus we may find some $x$ for which $\T_t(x)$ oscillates
(for instance an eigenvector of $\Ll$ with purely imaginary
eigenvalue). This can not happen if $\FT=\nT$.

\section{The structure of QMS with atomic deco\-her\-ence-free subalgebra}
\label{sect:struct-QMS}

In this section we prove our main results on the structure of
QMS with atomic decoherence-free subalgebra. The starting
point of our analysis is Proposition \ref{pi-inv}.
Since the central projections $p_i$ in (\ref{eq:NT-decomp}) are
fixed points for $\T_t$, by  Lemma \ref{lem:proj-FT}, we have
$\Ll^n(p_i x p_i) = p_i\Ll^n(x) p_i$, for all $n\ge 0$ and so
\begin{equation}\label{eq:Tt(pixpi)}
\T_t(p_i xp_i) = \T_t(p_i) \T_t(x) \T_t(p_i) =p_i \T_t(x) p_i
\end{equation}
for each $x\in\nT$ and each factor $p_i\nT p_i$ is $\T_t$-invariant.

Moreover, it would not be difficult to see that the
decoherence-free subalgebra of the restriction of $\T$ to
bounded operators on $p_i\h$ is $p_i\nT p_i$. Thus,
we begin by considering the case where $\nT$ is a
type I factor and investigate the implications on
the structure of $\T$.

We recall that, by well known results on the structure of
type I factors (see e.g. Jones \cite{jones}, Theorem 4.2.1),
in this case, there exist two Hilbert  spaces $\kk$ and $\mm$ and
a unitary operator $U:\h\to\kk\otimes\mm$ such that
\begin{equation}\label{eq:typeI-struct}
U\nT U^* = \mathcal{B}(\kk)\otimes\unit_{\mm}.
\end{equation}
Exploiting this structure we can prove our first result.

\begin{theorem}\label{th:NT-factor}
Suppose that $\nT$ is a type I factor, and let $\kk$, $\mm$ be
Hilbert spaces and $U:\h\to\kk\otimes\mm$ a unitary operator
satisfying (\ref{eq:typeI-struct}). Then:
\begin{enumerate}
\item for every GKSL representation (\ref{eq:GKSL}) of the
generator $\Ll$ by means of operators $L_\ell, H$, we have
\[
UL_\ell U^* = \unit_{\kk}\otimes M_\ell \quad\forall\,\ell\ge 1,
\qquad UHU^* = K\otimes \unit_{\mm}
+ \unit_{\kk}\otimes M_0,
\]
where $M_\ell$ are operators on $\mm$ such that the series
$\sum_{\ell}M_\ell^* M_\ell$ is strongly convergent and
$K$ (resp. $M_0$) is a self-adjoint operator on $\kk$
(resp. $\mm$),
\item  defining the GKSL generators $\Ll^k$ on
$\mathcal{B}(\kk)$ and $\Ll^\mm$ on $\mathcal{B}(\mm)$ by
\begin{eqnarray}
\label{def-Lk}\Ll^\kk(a)
& = & \mi\left[K,a\right],  \\
\label{def-Lm}\Ll^\mm(y)
&=&\mi\left[M_0,y\right]  \nonumber\\
&-&\frac{1}{2}\sum_{\ell\ge 1}\left(M_\ell^*M_\ell\, y
-2 M_\ell^*y M_\ell + y\, M_\ell^*M_\ell \right)
\end{eqnarray}
the QMSs $\T^\kk$ on $\mathcal{B}(\kk)$  generated by
$\Ll^\kk$ and $\T^\mm$ on $\mathcal{B}(\mm)$ generated
by $\Ll^\mm$ satisfy
$U\T_t(x)U^*=(\T_t^\kk\otimes\T_t^\mm)(UxU^*)$
for all $x\in\B$,
\item  we have $\T_t^\kk(a)=
\hbox{\rm e}^{\mi t K}a\,\hbox{\rm e}^{-\mi t K}$
for all $a\in\mathcal{B}(\kk)$, $t\ge 0$; moreover,
$\mathcal{N}(\mathcal{T}^\kk)=\mathcal{B}(\kk)$ and
$\mathcal{N}(\mathcal{T}^\mm)=\mathbb{C} \unit_\mm$.
\end{enumerate}
\end{theorem}

\noindent{\bf Proof.}
Let $L_\ell, H$ be the operators of a GKSL representation
(\ref{eq:GKSL}) of the generator $\Ll$. Since $\nT$ is
contained in the commutant of $L_\ell$ and $L_\ell^*$ by
Propostion \ref{prop:NT-comm}, it follows that $UL_\ell U^*$
and $UL_\ell^* U^*$ belong to the commutant of
$\mathcal{B}(\kk)\otimes\unit_{\mm}$ and so they
are operators of the form $\unit_\kk\otimes M_\ell$
and $\unit_\kk\otimes M_\ell^*$ for some bounded
operator $M_\ell$ on $\mm$.
The series $\sum_{\ell\ge 1}M_\ell^*M_\ell$ is strongly
convergent on $\mm$ because, if we fix a  vector $u\in\kk$,
then, for each vector $v\in\mm$ we have
\begin{eqnarray*}
u\otimes \left(\sum_{1\le \ell \le n} M_\ell^* M_\ell\, v\right)
& = & \sum_{1\le \ell \le n} \left(\unit_\kk \otimes M_\ell \right)^*
\left(\unit_\kk \otimes M_\ell \right) (u\otimes v) \\
& = &
U \left(\sum_{1\le \ell \le n} L_\ell^*L_\ell \right)
U^* (u\otimes v)
\end{eqnarray*}
for all $n\ge 1$, and the series $\sum_\ell L_\ell^* L_\ell$ is
strongly convergent on $\h$.

We now turn to $UHU^*$. For any $x\in\nT$ we have $U x U^*
= x_0\otimes \unit_{\mm}$   with $x_0\in \mathcal{B}(\kk)$
and $\T_t(x)=\mathrm{e}^{\mi tH} x\, \mathrm{e}^{-\mi tH} $,
so that
\[
U \T_t( U^*(x_0\otimes\unit_{\mm})U) U^*
=(U \mathrm{e}^{\mi tH} U^*)
\left( x_0\otimes\unit_{\mm}\right)
(U \mathrm{e}^{\mi tH} U^*)^*.
\]
By the $\T_t$-invariance of $\nT$, the right-hand side
has the form
\[
(W_t\otimes \unit_{\mm})(x_0\otimes\unit_{\mm})
(W_t\otimes \unit_{\mm})^*
\]
for a one-parameter group $(W_t)_{t\in\mathbb{R}}$ of
unitary operators on $\kk$. Thus, by defining
$V_t=U \mathrm{e}^{\mi tH} U^*$, we have
\[
V_t (x_0\otimes\unit_{\mm}) V_t^*
= (W_t\otimes \unit_{\mm})
(x_0\otimes\unit_{\mm})
(W_t\otimes \unit_{\mm})^*
\]
namely, for all $x_0\in\mathcal{B}(\kk)$,
\[
\left((W_t\otimes \unit_{\mm})^*V_t\right)
(x_0\otimes\unit_{\mm})
= (x_0\otimes\unit_{\mm})
\left((W_t\otimes \unit_{\mm})^*V_t\right).
\]
It follows that $(W_t\otimes \unit_{\mm})^*V_t$
must be of the form $\unit_{\kk}\otimes R_t$ with
unitaries $R_t$ on $\mm$ and, by the group property
of $(V_t)_{t\in\mathbb{R}}$, also $(R_t)_{t\in\mathbb{R}}$
must be a group.
Denoting $\mi K$ and $\mi M_0$  the generators of the
unitary groups $(W_t)_{t\in\mathbb{R}}$ and  $(R_t)_{t\in\mathbb{R}}$ respectively, we find
\begin{equation}\label{eq:UHUstar}
UH U^* = K\otimes \unit_{\mm}
+ \unit_{\kk}\otimes M_0.
\end{equation}
This proves 1.

To prove 2, note first that $\Ll^\kk$ and $\Ll^\mm$
generate QMSs $\T^\kk$ and $\T^\mm$ and
\[
U\Ll(U^*(a\otimes y)U)U^* =
\Ll^\kk(a) \otimes y + a \otimes\Ll^\mm(y),
\]
for all $a\in\mathcal{B}(\kk)$, $y\in\mathcal{B}(\mm)$,
so that, by the weak$^*$ density of the linear span of
operators $a\otimes y$ in $\B$, the QMSs
$(U\T_t(U^*\cdot U) U^*)_{t\ge 0}$ and
$(\T^\kk_t\otimes \T^\mm_t)_{t\ge 0}$ have the
same generator.

 Clearly $\T^\kk_t(a)= \hbox{\rm e}^{\mi t K} a\,
\hbox{\rm e}^{\mi t K}$ for all $t\ge 0$, and so
$\mathcal{N}(\T^\kk)= \mathcal{B}(\kk)$. Moreover,
if $p$ is a projection in $\mathcal{N}(\T^\mm)$, then,
by Proposition \ref{prop:NT-comm} of
the decoherence-free subalgebra, $p$ commutes with
all iterated commutators $\delta^n_{M_0}(M_\ell)$,
$\delta^n_{M_0}(M_\ell^*)$ ($n\ge 0, \ell\ge 1$).
Thus, recalling (\ref{eq:UHUstar}), $\unit_\kk\otimes p$
commutes with all iterated commutators
\[
\delta^n_{UHU^*}(\unit_\kk\otimes M_\ell)
=   U \delta^n_{H}(L_\ell) U^*\qquad
\delta^n_{UHU^*}(\unit_\kk\otimes M_\ell^*)
=  U \delta^n_{H}(L_\ell^*) U^*,
\]
i.e. it belongs to $U\nT U^*=\mathcal{B}(\kk)
\otimes\unit_{\mm}$ and so $p=\unit_\mm$.

This proves 3.
\hfill $\square$ \medskip

Theorem \ref{th:NT-factor} shows that maps $\T_t$
factorise as the composition of the commuting maps
$\T_t^\kk \otimes I_{\mathcal{B}(\mm)}$ and
$I_{\mathcal{B}(\kk)} \otimes \T_t^\mm$
where $I_{\mathcal{B}(\mm)}$ (resp. $I_{\mathcal{B}(\kk)}$)
is the identity map on $\mathcal{B}(\mm)$
(resp. $\mathcal{B}(\kk)$).
The former is the decoherence-free factor and the
latter is the decoherence-affected factor by item 3.
The generator $\Ll$ of $\T$ is the sum of two commuting
generators $I_{\mathcal{B}(\kk)} \otimes \Ll^\mm$ and
$\Ll^\kk\otimes I_{\mathcal{B}(\mm)}
=\mi\left[K\otimes\unit_{\mm},\cdot\,\right]$. This result can be
be interpreted as the independence of the decoherence-free
(noiseless) and the noisy part of the system.

\begin{remark}\label{rem:structure-atomic}
{\rm
If $\nT$ is an atomic algebra, by Proposition
\ref{atomic}, we can find a family $(p_i)_{i\in I}$
of mutually orthogonal projections which are
minimal in $\Z\nT$ such that $\sum_{i\in I}p_i=\unit$
and satisfying (\ref{eq:NT-decomp}).  Moreover,
each $p_i\nT p_i$ a type I factor acting on the Hilbert
space $p_i\h$; thus, there exist two countable sequences
of Hilbert spaces $(\kk_i)_{i\in I}$, $(\mm_i)_{i\in I}$,
 and unitary operators $U_i:p_i\h\to\kk_i\otimes\mm_i$ such that
\begin{equation}\label{NTi}
U_ip_i\nT p_iU_i^*
=\mathcal{B}(\kk_i)\otimes\unit_{\mm_i},\qquad
U_ip_i\B p_iU_i^*=\mathcal{B}(\kk_i\otimes\mm_i).
\end{equation}
Therefore, defining $U=\oplus_{i\in I} U_i$, we obtain a unitary
operator $U:\h\to \oplus_{i\in I}(\kk_i\otimes \mm_i)$ such that
\begin{equation}\label{eq:UnT}
U\nT U^*=\oplus_{i\in I}\left(\mathcal{B}(\kk_i)\otimes\unit_{\mm_i}\right).
\end{equation}
}
\end{remark}

We now establish our main result using the structure of $\nT$
given by (\ref{eq:UnT}).

\begin{theorem}\label{th:main}
Suppose that $\nT$ is an atomic algebra and let $(\kk_i)_{i\in I}$,  $(\mm_i)_{i\in I}$
be two countable sequences of Hilbert spaces and $U=\oplus_{i\in I} U_i$ be a unitary
operator associated with a family $(p_i)_{i\in I}$ as in Remark \ref{rem:structure-atomic}.
Then:
\begin{enumerate}
\item for every GSKL representation of $\Ll$ by means of operators
$H,(L_\ell)_{\ell\ge 1}$, we have
\[
UL_\ell U^*=\oplus_{i\in I}
\left(\unit_{\kk_i}\otimes M_\ell^{(i)}\right)
\]
for a collection $(M_\ell^{(i)})_{\ell\geq 1}$ of operators in
$\mathcal{B}(\mm_i)$, such that the series $\sum_{\ell\ge 1}M_\ell^{(i)*}M_\ell^{(i)}$
strongly convergent for all $i\in I$, and
\[
UHU^*=\oplus_{i\in I}\left(K_i\otimes\unit_{\mm_i}
+\unit_{\kk_i}\otimes M_0^{(i)}\right)
\]
for  self-adjoint operators $K_i\in\mathcal{B}(\kk_i)$ and $M_0^{(i)}\in\mathcal{B}(\mm_i)$,
$i\in I$,
\item  defining on the algebra $\mathcal{B}\left(\oplus_{i\in I}
\left(\kk_i\otimes\mm_i\right)\right)$
\begin{equation}\label{eq:L-da-df}
\dfL = \mi\left[\oplus_{i\in I}(K_i\otimes\unit_{\mm_i}),\cdot\,\right]
\end{equation}  and $\dL$ as the Lindblad operator given by
$$
\{\oplus_{i\in I}\left(\unit_{\kk_i}\otimes M_\ell^{(i)}\right),
\oplus_{i\in I}(\unit_{\kk_i}\otimes M_0^{(i)})\,|\,l\geq 1\},
$$
we find the commuting generators $\dfL$ and $\dL$ of two commuting QMSs
$\dfT$ (the decoherence-free semigroup) and $\dT$ (the decoherence-affected
semigroup)  such that $\wT_t=\dT_t\circ\dfT_t=\dfT_t\circ\dT_t$, where
$\wT$ is the QMS defined by
\begin{equation}\label{Ttilde}
\wT_t(UxU^*)=U\T_t(x)U^*\qquad\forall\,x\in\B.
\end{equation}
In particular, we have $$\dL(x) =\oplus_{i\in I}
\left(I_{\mathcal{B}(\kk_i)}\otimes\Ll^{\mm_i}\right)(x)=\oplus_{i\in I} \left(a_i\otimes\Ll^{\mm_i}(y_i)\right)$$ for all $x=\oplus_{i\in I} (a_i\otimes y_i)$
with $a\in\mathcal{B}(\kk_i)$ and $y\in\mathcal{B}(\mm_i)$,
where $\Ll^{\mm_i}$ is given by (\ref{def-Lm}),
\item the action of $\dfT$ is explicitly given by
$\dfT_t(x)=\hbox{\rm e}^{\mi t K }x \hbox{\rm e}^{-\mi t K }$ for all $x\in\mathcal{B}\left(\oplus_{i\in I}
\left(\kk_i\otimes\mm_i\right)\right)$,
where $K$ is the self-adjoint operator $\oplus_{i\in I} (K_i\otimes\unit_{\mm_i})$; moreover
$\mathcal{N}(\dfT)= \mathcal{B}\left(\oplus_{i\in I}
\left(\kk_i\otimes\mm_i\right)\right)$ and
$\mathcal{N}(\dT)=U\nT U^*$.
\end{enumerate}
\end{theorem}

\noindent{\bf Proof.}
Note that, since each $p_i$ is a fixed point for $\T_t$, by
Proposition \ref{pi-inv},  the algebra $p_i\B p_i=\mathcal{B}(p_i\h)$
is preserved by the action of every map $\T_t$, and so we can
consider the restriction of $\T$ to this algebra, getting a QMS
on $\mathcal{B}(p_i\h)$ denoted by $\T^{(i)}$.
Since, for all $x\in\nT$, by Proposition \ref{prop-struct-NT}
and Lemma \ref{lem:proj-FT} we have $\T_t(p_i x^* p_i x p_i)
=p_i\T_t(x^*p_i x)p_i =p_i \T_t(x^*) \T_t(p_i x)p_i
=p_i \T_t(x^*) p_i\T_t(x)p_i = \T_t(p_i x^*p_i)
\T_t(p_i xp_i)$, namely $p_i x p_i \in\nT^{(i)}$, it is easy to see
that the decoherence-free subalgebra $\mathcal{N}(\T^{(i)})$
of $\T^{(i)}$ is exactly $p_i\nT p_i$ . Moreover, given a
GSKL representation of $\Ll$ by means of operators
$H,(L_\ell)_{\ell\ge 1}$, since $p_i\in\nT$ commutes with every
$L_\ell$ by  Proposition \ref{prop:NT-comm},
and consequently also with $H$ (being $p_i$ a fixed
point),  the operators $p_iHp_i,(p_iL_\ell p_i)_{\ell\ge 1}$ provide
a GSKL representation of  the generator $\Ll^{(i)}$ of $\T^{(i)}$.
Therefore, applying Theorem \ref{th:NT-factor} to $\T^{(i)}$, we get
\[
U_ip_iL_\ell p_iU_i^*=\unit_{\kk_i}\otimes M_\ell^{(i)}\qquad
U_ip_iHp_iU_i^*
=K_i\otimes\unit_{\mm_i}+\unit_{\kk_i}\otimes M_0^{(i)}
\]
for some operators $K_i=K_i^*$ in $\mathcal{B}(\kk_i)$ and $M_0^{(i)}=(M_0^{(i)})^*,
(M_\ell^{(i)})_{\ell \geq 1}$ in $\mathcal{B}(\mm_i)$. Since $U=\oplus_{i\in I}U_i$, item $1$ is proved.

The claim 2 follows by the same argument of Theorem
\ref{th:NT-factor} claim 2.

The explicit action of $\dfT$ is also clear, and so
$\mathcal{N}(\dfT)= \mathcal{B}\left(\oplus_{i\in I}
\left(\kk_i\otimes\mm_i\right)\right)$.

Finally, by Proposition \ref{prop:NT-comm}, an operator $x$ is in
$\mathcal{N}(\dT)$  if and only if it commutes with all
iterated commutators
\begin{eqnarray*}
\delta_{\oplus_{i\in I}\left(\unit_{\kk_i}\otimes
M_0^{(i)}\right)}^{n}
\left(\oplus_{i\in I}
\left(\unit_{\kk_i}\otimes M_\ell^{(i)}\right)\right)
 & = &  \delta_{UHU^*}^n\kern-2truept
\left(\oplus_{i\in I}\kern-2truept
\left(\unit_{\kk_i}\otimes M_\ell^{(i)}\right)\right) \\
 & = &
 U\delta_{H}^n(L_\ell) U^*,  \\
\delta_{\oplus_{i\in I}\left(\unit_{\kk_i}\otimes
M_0^{(i)}\right)}^{n}\kern-2truept
\left(\oplus_{i\in I}\kern-2truept
\left(\unit_{\kk_i}\otimes M_\ell^{(i)*}\right)
\right)
& = &  \delta_{UHU^*}^{n}\kern-2truept
\left(\oplus_{i\in I}\kern-2truept
\left(\unit_{\kk_i}\otimes M_\ell^{(i)*}\right)
\right) \\
& = & U\delta_{H}^{n}(L_\ell^*) U^*,
\end{eqnarray*}
because all $K_j\otimes\unit_{\mm_j}$  and
$\unit_{\kk_i}\otimes M_\ell^{(i)}$ commute,
and so $x\in U \nT U^*$.
\par\null \hfill $\square$ \medskip

\begin{remark}\label{structure-i}\rm
Note that, in particular, the   central projection $p_i$
is minimal in $\nT$ if and only if $\kk_i$ is a one-dimensional
space, i.e. $U_ip_i\nT p_iU_i^*=\mathbb{C}\unit_{\mm_i}$.
Moreover, defining the QMS $\wT$ as in (\ref{Ttilde}),
we have
\[
\wT_t(a\otimes b)
=\hbox{\rm e}^{\mi tK_i}a
\hbox{\rm e}^{-\mi tK_i}\otimes\T_t^{\mm_i}(b)
\]
for all $a\in\mathcal{B}(\kk_i),\ b\in\mathcal{B}(\mm_i),\ i\in I,\ t\geq 0$, where $\T^{\mm_i}$ is the QMS on $\mathcal{B}(\mm_i)$ generated by $\Ll^{\mm_i}$. Finally, $\mathcal{N}(\T^{\mm_i})=\mathbb{C}\unit_{\mm_i}$ for all $i\in I$.
\end{remark}

Theorem \ref{th:main} also provides a constructive method
for finding the decoherence-free part of a quantum Markovian
evolution starting from the decomposition (\ref{eq:NT-decomp}).
The following proposition turns out to be useful when
we want to identify $\nT$.

\section{Structure of normal invariant states}

In this section we give a complete description of invariant states
of a QMS with atomic decoherence-free subalgebra.  We omit
the word normal in order to simplify the terminology since we
are interested only in normal states; moreover states will be
often identified with their densities.

We begin by recalling some well-known properties of  invariant states.
The \emph{support projection} $s(\rho)$ of a state $\rho$ is defined
as the orthogonal projection onto its range. More precisely,
if $\rho=\sum_{j\in J}\lambda_j\ee{j}{j}$ with $(e_j)_j$
orthonormal vectors in $\h$ and $\lambda_j>0$ for all $j\in J$,
then $s(\rho)=\sum_{j\in J}\ee{j}{j}.$
In particular, $\rho$ is faithful if and only if $s(\rho)=\unit$.

The support projection $p$ of  an invariant state  $\rho$  is
\emph{subharmonic} (\cite{FFRRjmp02}) Theorem II.1,
\cite{U} Theorem 1) i.e. $\T_t(p)\ge p$ for all $t\ge 0$.
Useful properties of subharmonic projections are collected in the
following proposition (see e.g. \cite{FV82}).

\begin{proposition}\label{prop:subharm-proj}
Let $p\in\B$ be a suhbarmonic projection. Then:
\begin{enumerate}
\item for all state $\sigma$ with $s(\sigma)\le p$,
the support of the normal state $\T_{*t}(\sigma)$ also
satisfies $s(\T_{*t}(\sigma))\le p$ for all $t\ge 0$,
\item $p\T_t(pxp)p = p\T_t(x)p $ for all $x\in\B$, $t\ge 0$
\item the one-parameter family of linear maps
$(\T_{t}^p)_{t\ge 0}$ defined by $\T_{t}^p(x)=
p\T_t(x)p$ for $x\in p\B p$ is a QMS on
$p\B p$, called the \emph{reduced} QMS,
\item $p$ is harmonic, i.e. $\T_t(p)=p$
for all $t\ge 0$, if and only if it belongs to the commutant
$\{L_k, L_k^*, H\,\mid\, k\geq 1\}^\prime$; in this case if
$\rho$ is a $\T$-invariant state such that $\tr{\rho p}\neq 0$, then
\begin{equation}\label{rhop}
\rho_p:={p\rho p}/{\tr{\rho p}}
\end{equation}
is an invariant state for the reduced QMS $\T^p$;
moreover, if $\rho$ is faithful, then $\rho_p$ is faithful
on $p\B p$ (i.e. $s(\rho_p)=p$).
\end{enumerate}
\end{proposition}

\noindent{\bf Proof.}
1. If $p^\perp$ is the orthogonal projection $\unit-p$,
for all $t\ge 0$ we find $0\le \tr{\T_{*t}(\sigma) p^\perp}
=\tr{\sigma \T_t(p^\perp)}\le \tr{\sigma p^\perp}=0$.
It follows that $p^\perp \T_{*t}(\sigma) p^\perp =0$
and so, by positivity of $\T_{*t}(\sigma)$,
we have $\T_{*t}(\sigma)=p\T_{*t}(\sigma)p$.

2. Let $x$ be a positive operator in $\B$.
Every state $\omega$ with support smaller than $p$  satisfies
$\omega= p\omega=\omega p$, therefore we have
$\tr{\omega p \T_t(pxp) p} =\tr{\T_{*t}(\omega) pxp} $.
Now, since also the support of $\T_{*t}(\omega)$ is smaller
than $p$, we find
\[
\tr{\omega p \T_t(pxp) p} = \tr{\T_{*t}(\omega) x}
= \tr{\omega \T_t(x)} =\tr{\omega p \T_t(x) p}
\]
and the conclusion follows.

3. For all $x\in p\B p$ and $t,s\ge 0$ we have from 2
\[
\T^p_{t+s}(x)  
=p\T_t\left(\T_s(x)\right)p
=p\T_t\left(p\T_s(x)p\right)p
=p\T_t\left(\T^p_s(x)\right)p
=\T_t^p\left(\T_s^p(x)\right).
\]
Moreover, since $p$ is subharmonic  and smaller than $\unit$,
we have also $\T_t^p(p) = p$.
Complete positivity and continuity properties are immediate.

4.  The first part of the claim follows from Lemma \ref{lem:proj-FT}.

It is clear that $s(\rho_p)\leq p$.
Since $p$ commutes with each $L_k,\, L_k^*$ and with $H$, we have $\T_t(x)=\T_t(pxp)=p\T_t(x)p$ for all $x\in p\B p$ and $t\ge 0$. Hence, we find
\[
\tr{\rho_p\T_t^p(x)}
=\frac{\tr{\rho \,p\T_t(pxp)p}}{\tr{\rho p}}
=\frac{\tr{\rho \T_t(x)}}{\tr{\rho p}}
=\frac{\tr{\rho x}}{\tr{\rho p}}=\tr{\rho_p x}
\]
and so $\rho_p$ is a $\T^p$-invariant state.
Assume now that $\rho$ faithful. Given $x\in\B$
such that $pxp\geq 0$, the equality
$0=\tr{\rho_px}=\tr{\rho pxp}/{\tr{\rho p}}$
implies $pxp=0$ by the faithfulness of $\rho$. Hence,
$\rho_p$ is faithful on $p\B p$.
\hfill $\square$ \medskip

 We refer the interested reader to the recent paper
\cite{Hachicha} for additional information on the support
of states evolving under the action of a QMS.

\smallskip

If $(q_i)_{i\in I}$ is a collection of subharmonic projections,
the projection $p$ onto the linear span of subspaces $q_i\h$
is also subharmonic (\cite{U} Proposition 3). We can then
define the \emph{(fast) recurrent projection} $p_R$ as the
smallest projection in $\h$ containing the support of all
invariant states
\[
p_R:=\sup\{s(\sigma)\,\mid\,\sigma\ \mbox{invariant state}\}.
\]
Moreover, we can always find an invariant state having $p_R$
as support (see Theorem $4$ of \cite{U}). As a consequence,
the reduced QMS $\T^{p_R}$ on $p_R\B p_R=\mathcal{B}(p_R(\h))$
has a faithful invariant state.

Since this section is devoted to the description of invariant states,
in the sequel, we consider this reduced semigroup dropping the
exponent $p_R$ and {\it assuming
the existence of a faithful invariant state.}

As a consequence we have the following

\begin{proposition}\label{prop:off-diag-pi-pj}
Let $\T$ be a QMS with a faithful invariant state $\rho$ and let
$\nT$ be as in (\ref{eq:NT-decomp}) with $(p_i)_{i\in I}$ minimal
projections in the center of $\nT$. Then $p_i\sigma p_j=0$ for
all $i\not=j$ and for every invariant state $\sigma$.
\end {proposition}

\noindent{\bf Proof.} Since central projections $p_i,p_j$ are in
$\mathcal{F}(\T)$, for all $t\ge 0$ and $x\in\B$ we have
$\T_t(p_j x p_i) = p_j \T_t(x) p_i$ and also
$\T_{*t}(p_i \sigma p_j) = p_i \T_{*t}(\sigma) p_j$ for
all trace class operator $\sigma$. It follows that
\[
\tr{p_i \sigma p_j x}
=\tr{p_i \T_{*s}(\sigma ) p_j x}
=\tr{ \sigma  \T_{s}(p_j x p_i )}
=\tr{ \sigma  p_j \T_{s}(x ) p_i}
\]
for all invariant state $\sigma$ and $x\in\B$.
Integrating on $[0,t]$ and dividing by $t$ we find
\[
\tr{p_i\sigma p_j x}  = \tr{\sigma p_j
\left(t^{-1}\int_0^t\T_{s}(x ) ds\right) p_i},
\]
and, taking the limit as $t\to\infty$, by
Theorem \ref{th:mean-ergodic}, we have,
\[
\tr{p_i\sigma p_j x} = \tr{p_i\sigma p_j \mathcal{E}(x)}
\]
where $\mathcal{E}(x)\in \mathcal{F}(\T)$. Now, since $\FT$
is also contained in $\nT=\oplus_{i\in I} p_i \nT p_i$,  we get
$p_j \mathcal{E}(x)p_i=0$ for $i\neq j$ as well as
$\tr{p_i\sigma p_j x}=0$, and so $p_i\sigma p_j =0$ by the
arbitrarity of $x$.
\hfill $\square$ \medskip

Item $4$ of Proposition \ref{prop:subharm-proj} and Proposition
\ref{prop:off-diag-pi-pj} show that, for studying the structure
of invariant states, (with a unitary transformation as in
Theorem \ref{th:NT-factor}) we can restrict ourselves to
the case where we are given a QMS $\T$ with
$\nT=\mathcal{B}(\kk)\otimes \unit_{\mm}$
with a faithful invariant state $\rho$. In other words,
we can now identify $\wT$ and $\T$ and suppose that
$\nT$ is a type I factor.

Before we begin our study of this case, it will be useful to remind
ourselves of some properties of partial traces. We refer to
S. Attal's  lecture notes \cite{Attal} for proofs. Given two
Hilbert spaces $\kk$ and $\mm$, for every $f\in \mm$ we
define the bounded operator
\[
\left| f \right\rangle_\mm : \kk \to \kk\otimes \mm, \qquad
\left| f \right\rangle_\mm  e = e\otimes f
\]
with adjoint operator
\[
_\mm \left\langle f \right| : \kk\otimes \mm \to \kk, \qquad
_\mm \left\langle f \right| u\otimes v
=\left\langle f,v\right\rangle u.
\]
For a trace-class operator $\sigma$ on $\kk\otimes\mm$ the
partial trace of $\sigma$ with respect to $\mm$ is the trace-class
operator on $\kk$ defined by
\[
\ptr{\mm}{\sigma} = \sum_{n\ge 1}\ \null
_\mm \left\langle f_n \right| \sigma \left| f_n \right\rangle_\mm,
\]
where $(f_n)_{n\ge 1}$ is an orthonormal basis of $\mm$.
It can be shown that the above series is convergent with respect
to the trace norm and its sum does not depend on the choice
of the orthonormal basis of $\mm$.
Moreover, the partial trace $\ptr{\mm}{\sigma}$ is
the only  trace-class operator on $\kk$ satisfying
$\tr{\sigma a\otimes \unit_\mm}= \tr{\ptr{\mm}{\sigma} a}$
for all $a\in\mathcal{B}(\kk)$.

\begin{lemma}\label{lem:Tk-Tm}
Let $\T$ be a QMS on $\mathcal{B}({\kk\otimes\mm})$
with an invariant state $\rho$ such that
$\T_t(a\otimes b) = \T^\kk_t(a)\otimes\T_t^\mm(b)$ for
all $t\ge 0$, $a\in\mathcal{B}(\kk),b\in\mathcal{B}(\mm)$
where $\T^\kk$ and $\T^\mm$ are  QMS on
$\mathcal{B}(\kk)$ and $\mathcal{B}(\mm)$ respectively.
The partial trace $\ptr{\mm}{\rho}$ (resp. $\ptr{\kk}{\rho}$) is
an invariant state for the QMS $\T^\kk$ (resp. $\T^\mm$).
Furthermore, if $\rho$ is faithful, then also $\ptr{\mm}{\rho}$
and $\ptr{\kk}{\rho}$ are faithful.
\end{lemma}

\noindent{\bf Proof.} For all $a\in\mathcal{B}(\kk)$, by the properties
of the partial trace, we have
\[
\tr{\ptr{\mm}{\rho}\T^\kk_t(a)}
= \tr{\rho \left(\T^\kk_t(a)\otimes\unit_\mm\right)}
= \tr{\rho \T_t(a\otimes\unit_\mm)}
\]
so that, by the invariance of $\rho$,
\[
\tr{\ptr{\mm}{\rho}\T^\kk_t(a)}
= \tr{\rho (a\otimes\unit_\mm)}
= \tr{\ptr{\mm}{\rho}a}.
\]

This proves that $\ptr{\mm}{\rho}$ is an invariant state
for  the QMS $\T^\kk$. Clearly, we can prove that
$\ptr{\kk}{\rho}$ is an invariant state for the  $\T^\mm$
in the same way.

Finally, if $\rho$ is faithful on $\mathcal{B}({\kk\otimes\mm})$,
then also $\ptr{\mm}{\rho}$ is faithful on $\mathcal{B}(\mm)$
because, for all positive $b\in\mathcal{B}(\mm)$,
$\unit_\kk\otimes b$ is positive and we have
$\tr{\ptr{\mm}{\rho} b}=\tr{\rho (\unit_\kk\otimes b)}$.
We can check in the same way that $\ptr{\kk}{\rho}$ is faithful.
\hfill $\square$ \medskip

We now study invariant states for the QMS $\T^\mm$. We
begin by recalling the notion of irreducibility and highlighting
its relationship with the structure of $\nT$.

\begin{definition}
A QMS $\T$ on a von Neumann algebra $\M$ is said to be
\emph{irreducible} if there exist no non-trivial projection
$p\in\M$ satisfying $\T_t(p)\ge p$ for all $t\geq 0$.
\end{definition}

\begin{proposition}\label{prop:Tirr-NTtriv}
Assume that $\nT$ is atomic and there exists a faithful
invariant state $\rho$.
If $\T$ is irreducible, then both $\nT$
and $\mathcal{F}(\T)$ are trivial.
\end{proposition}

\noindent{\bf Proof.}
Since $\mathcal{F}(\T)$ is a von Neumann subalgebra of
$\B$, if it were non-trivial, it would contain a non-trivial projection
$p$ so that $\T_t(p)=p$ contradicting irreducibility. \\
As a consequence, by Proposition \ref{pi-inv}, the center of
$\nT$ is trivial, i.e. $\nT$ is a type I factor and we can
apply Theorem \ref{th:NT-factor}. Let $\kk$ and $K$,
be as in Theorem \ref{th:NT-factor}.
If $K$ is not a multiple of the identity operator on $\kk$,
considering a non-trivial projection  $p$ on $\kk$ commuting
with $K$, the operator $p\otimes \unit_{\mm}$ is a non-trivial
projection $p$  which is a fixed point for $\T$
contradicting irreducibility. Thus, since $K$ is a multiple of the
identity operator, for all $a\in\nT$, we have $\T_t(a)=a$
so that $\nT=\mathcal{F}(\T)$ is trivial.
\hfill $\square$ \medskip

We now exploit properties of irreducible QMS for characterising
invariant states of semigroups $\T^{\mm_i}$.

\begin{theorem}\label{th:T-mm-irr}
Let $\T$ be a QMS on $\mathcal{B}(\kk\otimes\mm)$ with a faithful
invariant state $\rho$ and $\nT=\mathcal{B}(\kk)\otimes\unit_\mm$.
Then the QMS $\T^{\mm}$ on $\mathcal{B}(\mm)$ is irreducible,
has a unique invariant state $\tau_\mm$ and, for all trace-class
operator $\eta$ on $\mm$, we have
\begin{equation}\label{eq:qms-conv}
w-\lim_{t\to \infty}\T^\mm_{*t}(\eta) = \tr{\eta}\tau_\mm
\end{equation}
\end{theorem}

\noindent{\bf Proof.} Let $p$ be a non-zero subharmonic
projection for $\T^\mm$, i.e. $\T^\mm_t(p)\ge p$
for all $t\ge 0$, then $\unit_\kk\otimes p$
is a subharmonic projection for $\T$. By the invariance of $\rho$
we have $\tr{\rho (\T_t(\unit_\kk\otimes p) - \unit_\kk\otimes p)}
=0$, and so $\T_t(\unit_\kk\otimes p) = \unit_\kk\otimes p$
since $\rho$ is faithful. This means that
\[
\unit_{\kk}\otimes p\in\mathcal{F}(\T)\subseteq
\mathcal{N}(\T)=\mathcal{B}(\kk)\otimes\unit_{\mm},
\]
 i.e. $p=\unit_{\mm}$. Thus $\T^\mm$ is irreducible.

Moreover, since $\rho$ is a faithful invariant state for $\T$, its partial trace $\ptr{\kk}{\rho}$ is a faithful invariant state for $\T^\mm$ by Lemma \ref{lem:Tk-Tm}, and so $\mathcal{N}(\T^\mm)$ is trivial thanks to Proposition \ref{prop:Tirr-NTtriv}. Therefore, $\mathcal{F}(\T^\mm)=\mathcal{N}(\T^\mm)=\mathbb{C}\unit_\mm$.
It follows then from Theorem \ref{th:FT=NT}  that
$w^*-\lim_{t\to\infty}\T^\mm_t(b)\in\mathcal{F}(\T^\mm)$
exists and is a multiple of the identity operator.
Taking the trace with respect to the invariant state
$\tau_\mm:=\ptr{\kk}{\rho}$  the limit is easily
shown to be $\tr{\tau_\mm b}$.
It follows that, for all trace-class operator $\eta$ on
$\mathcal{B}(\mm)$ and all  $b\in \mathcal{B}(\mm)$
we have then
\[
\lim_{t\to \infty}\tr{\T^\mm_{*t}(\eta)b}
= \tr{\eta}\tr{\tau_\mm b}
\]
and (\ref{eq:qms-conv}) is proved.
\hfill $\square$ \medskip

In the proof of our result on the structure of invariant states
we need the following

\begin{lemma}\label{lem:K-pp-spec}
Let $\alpha=(\alpha_t)_{t\ge 0}$ be a semigroup of
automorphisms of $\mathcal{B}(\kk)$ given by $\alpha_t(a) =
\hbox{\rm e}^{\mi t K} a \hbox{\rm e}^{-\mi t K}$
for some bounded self-adjoint operator $K$ on $\kk$.
If $\omega$ is a faithful normal invariant state for $\alpha$, then
$K$ has pure point spectrum.
\end{lemma}

\noindent{\bf Proof.}
Let $\omega=\sum_{j\ge 1} \omega_j  q_j$ be the spectral  decomposition of
$\omega$ with strictly positive eigenvalues
in decreasing order $\omega_1>\omega_2 > \dots $ and $q_j$
finite-dimensional mutually orthogonal projections
such that $\sum_{j\geq 1}q_j=\unit$.
Clearly,  $\hbox{\rm e}^{-\mi t K} \omega\,
\hbox{\rm e}^{\mi t K}=\omega$ since $\omega$ is
an invariant state and $\hbox{\rm e}^{-\mi t K} \omega^n\,
\hbox{\rm e}^{\mi t K}=\omega^n$ by the homomorphism
property for all $n\ge 1$, and so
\begin{equation}\label{eq:omegan}
\sum_{j\ge 1}\omega_j^n\,
\hbox{\rm e}^{-\mi t K} q_j\, \hbox{\rm e}^{\mi t K}
=\sum_{j\ge 1} \omega_j^n\, q_j.
\end{equation}
Dividing both sides by $\omega_1^n>0$ and taking the
limit as $n\to\infty$, we find
$ \hbox{\rm e}^{-\mi t K} q_1 \hbox{\rm e}^{\mi t K} = q_1$
for all $t\ge 0$ and so $q_1$ commutes with $K$. Removing
the term $j=1$ in (\ref{eq:omegan}) we can prove by the
same argument that $q_2$ commutes with $K$ and so
on recursively. It follows that $K=\sum_j q_j K q_j$. Clearly,
each $q_jKq_j$ is a self-adjoint operator on the finite-dimensional
space $q_j\h$, and so we can find an orthonormal
basis of eigenvectors of $K$ of each of these subspaces.
Vectors of these orthonormal bases form an orthonormal basis
of $\h$ by the faithfulness of $\omega$.
\hfill $\square$ \medskip

We can now  prove the main theorem characterising the
structure of  $\T$-invariant states.

\begin{theorem}\label{th:inv-stat}
Let $\T$ be a QMS on $\mathcal{B}(\kk\otimes\mm)$ with a faithful
invariant state $\rho$ and $\nT=\mathcal{B}(\kk)\otimes\unit_\mm$
and let $\tau_\mm$ be the unique invariant state of the partially
traced semigroup $\T^\mm$.
If $\eta$ is a $\T$-invariant state, then
\begin{equation}\label{eq:form-eta}
\eta = \sigma \otimes \tau_\mm
\end{equation}
where $\sigma$ is a state on $\mathcal{B}(\kk)$ whose
density  commutes with $K$.
\end{theorem}

\noindent{\bf Proof.} By Lemma \ref{lem:K-pp-spec} we can find
an orthonormal basis $(e_j)_{j\ge 1}$ of eigenvectors
of $K$ so that $Ke_j = \kappa_je_j$ for some $\kappa_j
\in \mathbb{R}$. Moreover, if $\eta$ is an invariant state,
we can define trace-class operators on $\mm$  by
products of bounded and trace-class operators,  as
$\eta_{jk}= \null_{\kk}\langle e_j |\eta | e_k\rangle_\kk$
so that
\[
\eta=\sum_{j,k\ge 1} |e_j\rangle\langle e_k|
\otimes \eta_{jk}.
\]
By Theorem \ref{th:NT-factor}
\[
\eta = \T_{*t}(\eta)
=\sum_{j,k\ge 1} \hbox{\rm e}^{\mi (\kappa_k-\kappa_j)t}
|e_j\rangle\langle e_k| \otimes \T^\mm_{*t}(\eta_{jk})
\]
and so, by the linear independence of rank one operators
$|e_j\rangle\langle e_k|$,
\[
\hbox{\rm e}^{\mi (\kappa_j-\kappa_k)t}\eta_{jk} =
\T^\mm_{*t}(\eta_{jk})
\]
for all $j,k$. Each operator $\T^\mm_{*t}(\eta_{jk})$ tends
to $\tr{\eta_{jk}}\tau_\mm$ as $t\to\infty$ (in the weak topology)
by Theorem \ref{th:T-mm-irr}. Thus, if $\kappa_j\not=\kappa_k$,
we find $\tr{\eta_{jk}}=0$, while, if $\kappa_j=\kappa_k$ we
have $\tr{\eta_{jk}} \tau_\mm = \eta_{jk}$.
It follows that
\[
\eta = \sum_{j,k} \left(\tr{\eta_{jk}} |e_j\rangle\langle e_k|\right)
\otimes \tau_\mm.
\]
Defining $\sigma:=\sum_{j,k} \tr{\eta_{jk}} |e_j\rangle\langle e_k|$, (\ref{eq:form-eta}) follows. Finally, a straightforward computation yields
\[
K\sigma - \sigma K =
\sum_{j,k} \left( \tr{\eta_{jk}}(\kappa_j-\kappa_k) \right)
 |e_j\rangle\langle e_k|=0,
\]
since $\tr{\eta_{jk}}=0$ for $\kappa_j\neq\kappa_k$, and so
$K$ commutes with $\sigma$.
\hfill $\square$ \medskip

If $\nT$ is not a type I factor, but it is atomic, then from Theorem
\ref{th:inv-stat} and Proposition \ref{prop:off-diag-pi-pj} we have
immediately the following

\begin{theorem}\label{th:form-eta}
Assume that $\nT$ is atomic and there exists a faithful $\T$- invariant
state.  Let $(p_i)_{i\in I}$, $(\kk_i)_{i\in I}$, $(\mm_i)_{i\in I}$, $(K_i)_{i\in I}$, $U:\h\to\oplus_{i\in I}\left(\kk_i\otimes\mm_i\right)$ be as
in Theorem \ref{th:main}. A  $\T$-invariant state $\eta$ can be
written in the form
\[
U\eta U^*=\sum_{i\in I}
\tr{\eta p_i} \sigma_i \otimes \tau_{\mm_i}
\]
where, for every $i\in I$,
\begin{enumerate}
\item $\tau_{\mm_i}$ is the unique $\T^{\mm_i}$-invariant state
which is also faithful,
\item $\sigma_i$ is a density  on $\kk_i$ commuting with
$K_i$.
\end{enumerate}
\end{theorem}

\begin{remark}\label{rem:BN}\rm
Under the conditions of Theorem \ref{th:form-eta},
all $K_i$ have pure point spectrum by Lemma
\ref{lem:K-pp-spec}. By considering the spectral decomposition
$K_i=\sum_{j} \kappa_j q_{ij}$ with $(q_{ij})_{j\in J_i}$
mutually orthogonal projections such that $\sum_{j\in J_i}q_{ij}
=\unit_{\kk_i}$ (and $\kappa_j\not=\kappa_{j^\prime}$
for $j\not= j^\prime$), we can write the unitary isomorphism
\[
\kk_i\otimes\mm_i
=(\oplus_{j\in J_i} q_{ij}\kk_i)\otimes\mm_i.
\]
Now every density $\sigma$ commuting with $K_i$ can be written
in the form $\sigma = \sum_{j\in J_i} \sigma_j $ where
$(\sigma_j)_{j\in J_i}$ is a collection of positive trace-class
operators on subspaces $q_{ij}\kk_i$  normalized by
$\sum_{j\in J_i}\tr{\sigma_{j}}=1$.
Clearly,  if the projection $q_{ij}$ is one-dimensional,
i.e. the eigenvalue $\kappa_j$ is simple, then $\sigma_j$ is
a scalar $r_j$, say, in $[0,1]$.
As a consequence, every invariant state supported in
$\kk_i\otimes\mm_i$ turns out to be written
(up to a unitary isomorphism) in the form
\[
\sigma\otimes\tau_{\mm_i}=
\sum_{j\in J_i} \sigma_{j}\otimes \tau_{\mm_i}
= \sum_{j\in J_i,\, \hbox{\rmsmall dim}(q_{ij})=1 }
 r_j \tau_{\mm_i}
+ \sum_{j\in J_i,\, \hbox{\rmsmall dim}(q_{ij})>1 }
\sigma_{j}\otimes \tau_{\mm_i}
\]
for positive constants $r_j$ and arbitrary trace-class operators  $\sigma_j$ on eigenspaces of $K_i$ with dimension strictly bigger
than $1$.

From Theorem \ref{th:form-eta} it follows that
each invariant state can be written as
\[
\sum_{k} c_k  \tau_{k}
+ \sum_{m} d_m \eta_m\otimes \tau_m
\]
where $c_k$  and $d_m$ are non-negative numbers,
$\sum_k c_k + \sum_m d_m= 1$, $\eta_m$ can be
any density matrix on an eigenspace of some $K_i$
with dimension strictly bigger than $1$, and $\tau_m$
is the unique invariant state of some $\T^{\mm_i}$.

The same result holds for any QMS with atomic decoherence-free
subalgebra $\mathcal{N}(\T^R)$ of the semigroup $\T^R$
reduced by the fast recurrent projection $p_R$.

This generalises the result proved by Baumgartner and Narnhofer
in \cite{BN} (Theorem 7) in the finite dimensional case.
\end{remark}

\section{Applications to decoherence}\label{sect:appl}

In this section we apply our results to the study of environment
induced decoherence  (\cite{BO,CSU2,O,O2}) and to the identification
of subsystems of an open quantum system which are not affected
by decoherence (\cite{AFR,LW,KL,TV}).

\subsection{Environment induced decoherence}

We say that there is \emph{environmental induced decoherence}
(EID) on the system described by $\T$ if there exists a
$\T_t$-invariant and $*$-invariant weak* closed subspace
$\M_2$ of $\B$ such that:
 \begin{itemize}
 \item[(EID1)] $\B=\nT\oplus\M_2$ with $\M_2\not=\{0\}$,
 \item[(EID2)] $w^*-\lim_{t\to\infty}\T_t(x)=0$ for all $x\in\M_2$.
 \end{itemize}
We refer to \cite{CSU2} Section 2 for a discussion of this concept.
As an application of our result on the structure of QMS we
will now give a sufficient condition for EID.

\begin{theorem}\label{th:EID}
Assume that $\nT$   is atomic and $\T$ possesses
a faithful normal invariant state. Then EID holds.
\end{theorem}

\noindent{\bf Proof.}
Since $\T$ possesses a faithful normal invariant state $\rho$, say,
it is enough to establish the existence of a normal
conditional expectation $\E$ onto $\mathcal{N}(\T)
=\oplus_{i\in I}\left(\mathcal{B}(\kk_i)\otimes\unit_{\mm_i}\right)$
which is compatible with $\rho$, i.e. such that $\rho\circ\E=\rho$
(Theorem 18 of \cite{CSU-new}).

Given $x\in\mathcal{B}(\h)$,  we write $x=\sum_{l,m}
p_l x p_m$ with minimal projections in the center of $\nT$
as in (\ref{eq:NT-decomp}), and identify each $p_l x p_m$
with a bounded operator $x_{lm}:\kk_m\otimes\mm_m\to\kk_l\otimes\mm_l$.
 Identifying an operator on $\kk_i\otimes\mm_i$ with its
extension  as the zero operator on the orthogonal subspace, we
then define
\[
\E:\mathcal{B}(\h)\to \nT,\qquad \E(x):=\sum_{i\in I}\E_i(x_{ii})
\]
with $\E_i:\mathcal{B}(\kk_i\otimes\mm_i)\to\mathcal{B}(\kk_i)\otimes\unit_{\mm_i}$ given by
\[
\E_i(a)=\sum_j\mbox{}_{\mm_i}\langle f_j\,|\,(\unit_{\kk_i}\otimes\tau_{\mm_i})a\,|\,f_j\rangle_{\mm_i}\otimes\unit_{\mm_i},
\]
for each $a\in \mathcal{B}(\kk_i\otimes\mm_i)$, where $\tau_{\mm_i}$
is the unique faithful invariant state for $\T^{\mm_i}$ and
$(f_j)_j$ is an orthonormal basis of $\mm_i$ diagonalizing $\tau_{\mm_i}$.
It is easy to see that every $\E_i$ is a positive normal map
such that $\E_i^2=\E_i$ and $\E_i(\unit_{\kk_i\otimes\mm_i})
=\unit_{\kk_i\otimes\mm_i}$, so that each $\E_i$ is a
normal conditional expectation onto
$\mathcal{B}(\kk_i)\otimes\unit_{\mm_i}$.
Consequently, $\E$ is a normal conditional expectation onto
$\nT=\oplus_{i\in I}
\left(\mathcal{B}(\kk_i)\otimes\unit_{\mm_i}\right)$.

Now, we have to show that $\E$ is compatible with $\rho$. First, note that, since $\rho$ is invariant, by Theorem \ref{th:form-eta}  we have $\rho=\sum_{i\in I}\sigma_i\otimes\tau_i$ for some
trace-class operators $\sigma_i$ on $\kk_i$ commuting with
$K_i$. Therefore, $\tr{\rho x}
=\sum_{i\in I}\tr{(\sigma_i\otimes\tau_i)x_{ii}}$,
with $x_{ii}=p_ixp_i$, for all $x\in\mathcal{B}(\h)$, and $\tr{\rho\E(x)}=\sum_{i\in I}\tr{\rho\E_i(x_{ii})}$, so that it is enough to prove that every $\E_i$ is compatible with $\sigma_i\otimes\tau_i$.

Now, for $a\in\mathcal{B}(\kk_i\otimes\mm_i)$ we easily compute
\begin{eqnarray*}
\tr{(\sigma_i\otimes\tau_i)\E_i(a)}
&=&\sum_j\tr{(\sigma_i\otimes\tau_i)(\mbox{}_{\mm_i}
\langle f_j\,|\,(\unit_{\kk_i}\otimes\tau_i)a\,|\,f_j\rangle_{\mm_i}\otimes\unit_{\mm_i})}\\
&=&\sum_j\tr{\sigma_i\,\mbox{}_{\mm_i}\langle f_j\,|\,(\unit_{\kk_i}\otimes\tau_i)a\,|\,f_j\rangle_{\mm_i}\otimes\tau_i}\\
&=&\sum_j\tr{\sigma_i\,\mbox{}_{\mm_i}\langle f_j\,|\, (\unit_{\kk_i}\otimes\tau_i)a\,|\,f_j\rangle_{\mm_i}}\\
& =&\tr{(\sigma_i\otimes\unit_{\mm_i})(\unit_{\kk_i}\otimes\tau_i)a}=\tr{(\sigma_i\otimes\tau_i)a},
\end{eqnarray*}
from which the required result follows.
\hfill $\square$ \medskip

\subsection{Decoherence-free subsystems and subspaces}

A quantum subsystem can be thought of intuitively as ``portion''
of the full system, whose states, in the simplest setting, faithfully
embody quantum information. More precisely, following Ticozzi
and Viola (\cite{TV} Definition 4), we call \emph{quantum subsystem}
of a system on $\h$ a system whose Hilbert space is a tensor factor $\h_\mathsf{s}$ of a subspace $\h_{\mathsf{sf}}$ of $\h$, i.e.
\begin{equation}\label{hqs}
\h=\h_{\mathsf{sf}}\oplus \h_\mathsf{r}
=(\h_\mathsf{s}\otimes\h_\mathsf{f})\oplus \h_\mathsf{r}
\end{equation}
for some factor $\h_\mathsf{f}$ and remainder space $\h_\mathsf{r}$.

\begin{definition}
Let $\h$ be decomposed as in (\ref{hqs}). We say that
$\h_{\mathsf{s}}$ supports a \emph{decoherence-free
(or a noiseless) subsystem} for some QMS $\T$ on $\B$ if
and only the evolution of a factorised initial state
$\rho=\rho_\mathsf{s}\otimes\rho_\mathsf{f}$, with
$\rho_{\mathsf{s}}$ state on $\mathcal{B}(\h_{\mathsf{s}})$
and $\rho_{\mathsf{f}}$ state on $\mathcal{B}(\h_{\mathsf{f}})$,
is given by
\[
\T_{*t}(\rho)=U_t \rho_{\mathsf{s}}U_t^* \otimes
\T^{\mathsf{f}}_{*t}(\rho_\mathsf{f})
\]
for $t\geq 0$, where $U_t$ is a unitary operator on $\h_{\mathsf{f}}$
and $\T^{\mathsf{f}}$ is a QMS on $\mathcal{B}(\h_\mathsf{f})$.
We say that $\h_{\mathsf{s}}$ supports a \emph{decoherence-free
subspace} if $\h_\mathsf{f}$ is one-dimensional, i.e.
$\h\simeq \h_\mathsf{s}\oplus\h_\mathsf{r}$.
\end{definition}

Note that the above definition of decoherence-free subspace is clearly
equivalent to the usual one (see \cite{LW} Definition 1).

Applying Theorem \ref{th:main} we can easily identify quantum
subsystems for a given QMS. Indeed, using the same notation
as in Theorem \ref{th:main}, we have the following

\begin{proposition}\label{nT-DFs}
If $\nT$ is an atomic algebra then:
\begin{enumerate}
\item  every subspace $\kk_i$  with factor $\mm_i$ and remainder
space $\oplus_{j\in I\setminus\{i\}}(\kk_j\otimes \mm_j)$, supports
a decoherence-free subsystem for $\T$,
\item every $\kk_i$ with ${\rm dim}\,\mm_i=1$ supports a decoherence-free subspace for $\T$; in particular, we have $M_\ell^{(i)}=\lambda_\ell^{(i)}\unit_{\mm_i}$ for all $\ell$ and $M_0^{(i)}=\alpha_i\unit_{\mm_i}$ for some $\lambda_\ell^{(i)},\,\alpha_i\in\mathbb{R}$,
\item  every subspace $\oplus_{j\in J} \kk_j$,
where $J\subseteq I_1:=\{ i\in I \mid\hbox{\rm dim}\,\mm_i=1, \lambda_\ell^{(i)}=\lambda_\ell\ \forall\,\ell \}$ (for some collection $(\lambda_\ell)_{\ell\geq 1}\subseteq\mathbb{R}$)
with trivial factor $\h_\mathsf{f}=\mathbb{C}$ and remainder space
$\oplus_{j\in I\setminus J}(\kk_j\otimes \mm_j)$,
supports a decoherence-free subspace for $\T$.
\end{enumerate}
\end{proposition}

\noindent{\bf Proof.} All the above can be proved straightforwardly applying
Theorem \ref{th:main}. We check, for instance, item $3$.

We have to show that
\begin{equation}\label{T-somma}
\T_{*t}(\kb{u}{u}\otimes \unit)
=\hbox{\rm e}^{-\mi t K} \kb{u}{u} \hbox{\rm e}^{\mi t K}
\otimes \unit
\end{equation}
for some self-adjoint operator $K$ on $\oplus_{j\in J} \kk_j$
and for all $u\in\oplus_{j\in J} \kk_j$.

So, let $u = \sum_{j\in J} u_j $ with $u_j\in \kk_j$ for $j\in J$,
and let $K_j$ be the self-adjoint operator on $\kk_j$ in
Theorem \ref{th:main}(2), identified with its standard ampliation
to $\h$. Theorem \ref{th:main} gives
\[
\T_{*t}(\kb{u_j}{u_j}\otimes \unit)
=\hbox{\rm e}^{-\mi t K_j} \kb{u_j}{u_j}
\hbox{\rm e}^{\mi t K_j}\otimes \unit.
\]
Now, given $j,h\in J$, we have
\begin{eqnarray*}
\Ll_{*}(\kb{u_j}{u_h}\otimes \unit)
&=&-\mi\left(\kb{(K_j+\alpha_j)u_j}{u_h}-\kb{u_j}{(K_h+\alpha_h)u_h}\right)\otimes \unit\\
&=&-\mi [(K_j+\alpha_j)\oplus(K_h+\alpha_h),\kb{u_j}{u_h}]
\otimes \unit,
\end{eqnarray*}
since the action of the dissipative part on $\kb{u_j}{u_h}$ is $0$
being $M_l^{(j)}=\lambda_\ell=M_\ell^{(h)}$ for $j,h\in J$ and
for every $\ell$. Therefore, we obtain
\begin{eqnarray*}
\Ll_*(\kb{u}{u})&=&\sum_{j,h\in J}\Ll_*(\kb{u_j}{u_h}) \\
& = &-\mi\sum_{j,h\in J}[(K_j+\alpha_j)\oplus(K_h+\alpha_h),\kb{u_j}{u_h}]\otimes \unit\\
&=&-\mi\sum_{j,h\in J}\left[\oplus_{l\in J}(K_l+\alpha_l),\kb{u_j}{u_h}\right]\otimes \unit\\
&=&-\mi \left[\oplus_{j\in J}(K_j+\alpha_j),\kb{u}{u}\right]
\otimes \unit
\end{eqnarray*}
and, consequently, equation (\ref{T-somma}) is satisfied with $K:=\oplus_{j\in J}(K_j+\alpha_j)$.
\hfill $\square$ \medskip

\section{Examples}
\subsection{Generic semigroups}
Take $\h=\ell^2(I)$ the Hilbert space of square-summable,
complex-valued sequences, with $I\subseteq\mathbb{N}$ a finite
or infinite set, and denote by $(e_n)_{n\in I}$
the canonical orthonormal basis of $\h$.

We consider a class of uniformly  continuous QMSs on $\B$
whose generators can be represented in the canonical GKSL form
\[
\Ll(x) = G^* x + \sum_{{j}
\neq m}L_{mj}^* x L_{mj} + xG
\]
where
\begin{eqnarray*}
 G = -\sum_{m\in {\mathbb N}} \left(\frac{\gamma_{mm}}{2}
 + i \kappa_m \right) |e_m\rangle \langle e_m|,
\qquad
L_{mj}=\sqrt{\gamma_{mj}}\,|e_j\rangle\langle e_m|,
\end{eqnarray*}
for $j\neq m$ with $\kappa_m\in\mathbb{R}$,
$\gamma_{mj}\geq 0$ for
every $m\neq j$,  $\gamma_{mm}:=-\sum_{j\neq m}\gamma_{mj}<+\infty$ for any $m$, and
\begin{equation}\label{L-bounded}
\sup_i|\kappa_i|<+\infty,\qquad\sup_i|\gamma_{ii}|<+\infty.\end{equation}
Note that $\Ll$ is bounded thanks to condition (\ref{L-bounded}),
and can be written in the form
\[
\Ll(x)=i[H,x]-\frac{1}{2}\sum_{j\neq m}\left(L_{jm}^*L_{jm}x-2L_{jm}^*xL_{jm}
+xL_{jm}^*L_{jm}\right)
\]
with $H=\sum_{j\in I}\kappa_j|e_j\rangle\langle e_j|$.

These semigroups, called \emph{generic}, were introduced by Accardi and Kozyrev in \cite{AK}; they arise in the stochastic limit of a open
quantum system with generic Hamiltonian, interacting with a  zero
mean, gauge invariant, Gaussian field (see also \cite{AFH,CFH}).

The restriction of $\Ll$ to the diagonal algebra $\D$, generated
by rank one projections $\ee{n}{n}$, is the
generator $\Gamma$ of a classical time continuous Markov
chain  $(X_t)_{t\geq 0}$ with states $I$ (see \cite{CFH}).
In particular, denoted by $\T$ the QMS generated by $\Ll$,
for every $x=\sum_{n\in I} f(n)\ee{n}{n}\in\D$, we have

\begin{eqnarray}\label{LsuD}
& & \Ll(x)=\sum_{n\in I}(\Gamma f)(n)\ee{n}{n},\ \
\Gamma f=\sum_{j\in I}\gamma_{nj}f(j)e_j,\\
\label{TsuD}
&  &\T_t(\ee{n}{n})=\sum_{k\in I}\PP
\left\{X_t=n\,|\,X_0=k\right\}\ee{k}{k}\quad  \forall\,n\in I.
\end{eqnarray}

For these semigroups, we recall the characterisation of the decoherence-free subalgebra
$\nT$ (see \cite{generic} Theorem 9 and \cite{CRSU} Theorem 26). Note first of all that
$[H,L_{mj}]=(\kappa_j-\kappa_m) L_{mj}$ and so
$\delta^{n}_H(L_{mj})=(\kappa_j-\kappa_m)^n L_{mj}$
for all $n\ge 0$. It follows immediately from Proposition
\ref{prop:NT-comm} that $\nT$ is the commutant of the set $\{L_{mj},L_{mj}^*\,\mid\, j,m\in I\}$,
i.e. the commutant of the set of rank-one
operators  $\{\,\ee{j}{m},\ee{m}{j}\,:\, j,m\in I,
\ \gamma_{jm}+\gamma_{mj}>0\,\}$. Clearly a self-adjoint
$x$ belongs to this commutant if and only if
$\left| xe_j\right\rangle\left\langle e_m\right|
= \left| e_j\right\rangle\left\langle x e_m\right|$, namely
$x e_j=\chi_j e_j$ and $x e_m = \chi_m e_m$ with
$\chi_j=\chi_m=\chi\in \mathbb{R}$. Moreover,
if $\gamma_{jm}\not=0$,
but there exists $k_1,\dots k_n$ in $I$ with $\gamma_{jk_1}
\gamma_{k_1 k_2} \cdots \gamma_{k_n m}\not=0$,
then $x e_{j}=\chi e_j, xe_{k_1}=\chi e_{k_1},
\dots, xe_m=\chi e_m$.

Thus, denoting by $\mathcal{C}_n$ $n\ge 1$, communication
classes of states $i$ such that $\gamma_{ij}+\gamma_{ji}>0$
for some $j\not=i$  with respect to the standard equivalence
relation on states associated with the rates matrix $Q$
obtained from $\Gamma$, for example, in the following way
$Q_{ij}=\gamma_{ij}+2^{-j}1_{\{\gamma_{ji}>0\}}
\ \mbox{for $i\neq j$}$,
$Q_{ii}=-\sum_{j\neq i}Q_{ij}$,
and denoting by
\[
\Iso:=\{i\in I:\,\gamma_{ij}=\gamma_{ji}=0\ \, \forall\,j\neq i\}
\]
the set of the isolated states of the Markov chain $(X_t)_t$,
we have the following result:

\begin{proposition}\label{prop:NT-generic}
The decoherence-free subalgebra of the generic QMS generated
by (\ref{L-bounded}) is the von Neumann subalgebra of  $\B$
generated by projections $p_n =\sum_{k\in \mathcal{C}_n}
\ee{k}{k}$ with $n\ge 1$, corresponding to the above
communication classes, and by rank-one operators $\ee{i}{j}$
with $ i,j\in \Iso$.
\end{proposition}

\noindent{\bf Proof.}
We want now to find a maximal family of mutually orthogonal minimal
projections in ${\Z\nT}$, in order to apply Theorem \ref{th:main}.

An operator $x\in\nT$ can be represented as $x_0 + \sum_{n\ge 1} \chi_n p_n$
with $x_0=p_0 xp_0\in \mathcal{B}(p_0\h)$ and $\chi_n\in \mathbb{C}$.
Therefore, an element $y=y_0 + \sum_{n\ge 1} z_n p_n$ in $\nT$, with
$y_0=p_0 yp_0\in\mathcal{B}(p_0\h)$  and $z_n\in \mathbb{C}$, commutes with
all $x\in\nT$ if and only if $y_0$ commutes with all operators
$x_0\in\mathcal{B}(p_0\h)$, namely $y_0 =z_0 p_0$ for some $z_0\in\mathbb{C}$.
Consequently, since every $p_n$ with $n\geq 1$ is minimal in $\nT$,
Remark \ref{structure-i} formula (\ref{NTi}), yields
\[
p_n\h\simeq\mm_n\quad\forall n\ge 1,\qquad p_0\h\simeq\mathcal\kk_0
\]
for some complex Hilbert spaces $(\mm_n)_{n\ge 1}$ and $\kk_0$, while Hilbert
spaces $\mm_0$ and $(\kk_n)_{n\ge 1}$ are one-dimensional. By Theorem \ref{th:main},
the decoherence-free and decoherence-affected semigroups are generated by
\[
\dfL = \mi [\,p_0 H p_0\, , \cdot \,] \qquad \dL = \Ll - \dfL.
\]
\par\noindent\hfill $\square$

\subsection{Circulant QMS}

In the previous example for each minimal central projection
$p_i$ either $\kk_i$ or $\mm_i$ is trivial, i.e. one-dimensional.
We now consider a paradigm case exhibiting non-trivial
$\kk_i$ and $\mm_i$  for the same index $i$.

Let $\h=\mathbb{C}^d$ ($d\ge 2$), let $n\in \{1,\dots, d-1\}$
and let $J$ the unitary circular shift defined by $J e_i = e_{i-1}$
(sum modulo $d$) with respect to an orthonormal basis of $\h$.
We consider the QMS on $\B =M_d(\mathbb{C})$ generated by
(\ref{eq:GKSL}) with
\[
L_1 = z_1 J^n, \qquad L_2 = z_2 J^{*n}=z_2 J^{-n}
\]
where $z_1,z_2$ are complex constants with
$z_1\cdot z_2\not=0$, $L_\ell = 0$ for $\ell\ge 2$. We begin
by considering $H=0$.
This is a circulant QMS of those studied by Bola$\tilde{\rm n}$os
and Quezada in \cite{BQ}.

Let $k=GCD(n,d)$ (Greatest Common Divisor) and let $d=km$.
The decoherence-free subalgebra $\nT=\{\, J^n, J^{n*}\}^\prime$
is  characterised as follows.

\begin{proposition}\label{prop:NT-circulant}
The algebra $\nT$ is
the commutant of  $\{ J^k, J^{-k}\}^\prime $.
\end{proposition}
\noindent{\bf Proof.}
The set of powers $hn$ (mod $d$) with $h\in\mathbb{Z}$
is $\{0,k,2k,\dots, (m-1)k\}$. Indeed, since $n k^{-1}$
and $dk^{-1}$ are coprime,  we can find integers $a,b$ such that
$ a n k^{-1} + b d k^{-1}=1$ i.e. $k=an+bd$,  and so
$k=an$ mod $d$.  Let $x\in\nT=\{J^n,J^{-n}\}^\prime$,
then $[x,J^k]=[x,J^{an}]=0$ and $[x,J^{-k}]=[x,J^{-an}]=0$,
thus $\nT\subseteq  \{ J^k, J^{-k}\}^\prime$. On the other
hand, since $n=h k$ for some natural number $h$, if
$[x,J^k]=0=[x,J^{-k}]$, by induction on $j$ we have also
$[x,J^{kj}]=0=[x,J^{-kj}]$ and so $[x,J^{n}]=[x,J^{ h k }]=0$,
$[x,J^{-n}]=[x,J^{-hk}]=0$, namely $  \{ J^k, J^{-k}\}^\prime
\subseteq\nT$.
\hfill $\square$ \medskip

\begin{proposition}
The center $\mathcal{Z}(\nT)$ of $\nT$ is the double commutant
$\{ J^k,J^{-k}\}^{\prime\prime}$ (the abelian $^*$-algebra
generated by $J^k$), namely the linear space $\mathsf{P}(J^k)$
of complex polynomials in $J^{kh}$ for $0\le h\le m-1$.
\end{proposition}

\noindent{\bf Proof.}
Clearly
\[
\mathcal{Z}(\nT)= \nT \cap  \nT^\prime
=\{ J^k, J^{-k}\}^\prime \cap \{ J^k, J^{-k}\}^{\prime\prime}.
\]
Note that $J^k,J^{-k}\in \{ J^k, J^{-k}\}^\prime$ therefore
the double commutant of $\{ J^k, J^{-k}\}$ is contained in
the commutant of $\{ J^k, J^{-k}\}$ and
$\mathcal{Z}(\nT)=\{ J^k, J^{-k}\}^{\prime\prime}$.
Since $J$ is a normal operator this is the algebra generated
by $J^k$. Indeed, by definition, all powers $J^{kh}$ are contained
in $\{ J^k,J^{-k}\}^{\prime\prime}$ because they commute with
every operator commuting with $J^k$ and $J^{-k}$ and so
$ \mathsf{P}(J^k)\subseteq \{ J^k,J^{-k}\}^{\prime\prime} $.
Conversely, $ \mathsf{P}(J^k)$ is a $^*$-algebra and contains
$J^{k}$ and $J^{-k}=J^{k(d-1)}$, therefore it coincides
with  $\{ J^k,J^{-k}\}^{\prime\prime} $.
Finally we can restrict exponents to $kh$ with
$0\le h\le m-1$ by the Cayley-Hamilton theorem.
\hfill $\square$ \medskip

We now identify minimal central projections. Since
the $^*$-algebra generated by the normal operator $J^k$
coincides with the vector space generated by its spectral projections,
these turn out to be minimal central projections in $\nT$.

Spectral projections of $J^k$ can be found explicitly from the
well-known spectral decomposition of the circulant matrix $J$.
Let $\omega := \hbox{\rm e}^{2\pi \mi /d}$ be a primitive
$d$-th root of unit; eigenvalues of $J$ and corresponding
eigenvectors are
\[
\lambda_j = \omega^j
\qquad v_j = \frac{1}{\sqrt{d}}(1,\omega^j,\dots,\omega^{j(d-1)}).
\]
for $j=0,\ldots,d-1$.  It follows that
\[
J   = \sum_{j=0}^{d-1} \omega^j \kb{v_j}{v_j}
\qquad \hbox{\rm and so} \qquad
J^k = \sum_{j=0}^{d-1} \omega^{jk} \kb{v_j}{v_j}.
\]

Now, writing $j\in\{0,\ldots,d-1\}$ as $j=mr+h$ with
$0\le h\le m -1 = d\, k^{-1}-1$, we find that $r$ must belong
to $\{0,\ldots,k-1\}$ and the eigenvalues of $J^k$ are
\[
\omega^{jk}=\hbox{\rm e}^{2\pi \mi (r+h/m)}
=\hbox{\rm e}^{2\pi \mi h/m},\qquad h=0,\ldots,m-1,
\]
since $d=km$. Moreover, for each $h=0,\ldots,m-1$, eigenvectors
of $\omega^{jk}=\hbox{\rm e}^{2\pi \mi (r+h/m)}$ are
vectors $v_{mr+h}$ with $r\in \{0,\ldots,k-1\}$.
As a result of these computations defining
\begin{equation}\label{eq:m-c-proj-circ}
p_{h} = \sum_{r=0}^{k-1}
\kb{v_{mr+h}}{v_{mr+h}},\qquad h=0,\ldots,m-1,
\end{equation}
we have
\begin{eqnarray*}
J^k   = \sum_{j=0}^{d-1} \omega^{jk} \kb{v_j}{v_j}
& = & \sum_{h=0}^{m-1} \sum_{r=0}^{k-1}
\omega^{(mr + h )k}\kb{v_{mr+h}}{v_{mr+h}} \\
& = & \sum_{h=0}^{m-1} \sum_{r=0}^{k-1}
\hbox{\rm e}^{2\pi\mi h/m}\kb{v_{mr+h}}{v_{mr+h}} \\
& = & \sum_{h=0}^{m-1}
\hbox{\rm e}^{2\pi\mi h/m} p_h.
\end{eqnarray*}
and $(p_h)_{0\le h\le m-1}$ is the collection of all
\emph{minimal central projections} in $\nT$.

We can now read off the factorisations of Theorems
\ref{th:NT-factor} and \ref{th:main}.
Let $(f_r)_{0\le r\le k-1}$ be an
orthonormal basis of $\mathbb{C}^k$. For each $h=0,\ldots,m-1$ define the unitary operator
\[
U_h:p_h\mathbb{C}^d\to \mathbb{C}^k,
\qquad
U_h v_{mr+h} = f_r,\quad r=0,\ldots,k-1,\]
so that we obtain the unitary $U:=\oplus_{h=0}^{m-1}U_h$
\[
U:\mathbb{C}^d\to\oplus_{h=0}^{m-1}\mathbb{C}^k
=\oplus_{h=0}^{m-1}\left(\kk_h\otimes\mm_h\right)
\]
with $\kk_h=\mathbb{C}^k, \mm_h=\mathbb{C}$ for
$h=0,\dots, m-1$.\\
It turns out that $U_hJ^kU_h^*f_r=\omega^{(mr+h)k}f_r=\hbox{\rm e}^{2\pi\mi h/m}f_r$ for all $r=0,\ldots,k-1$, i.e. $U_hJ^kU_h^*=S_h$ with $S_h:=\hbox{\rm e}^{2\pi\mi h/m}\unit_{\mathbb{C}^k}$. Hence, we have
$$U_hJ^nU_h^*=S_h^{n/k}=\hbox{\rm e}^{2\pi\mi hn/d}\unit_{\mathbb{C}^k}\quad\mbox{and}\quad UJ^nU^*=\oplus_{h=0}^{m-1}\left(\hbox{\rm e}^{2\pi\mi hn/d}\unit_{\mathbb{C}^k}\right).$$
Finally, since $U\nT U^*=\oplus_{h=0}^{m-1}\{U_hJ^nU_h^*,U_hJ^{-n}U_h^*\}^\prime$, we immediately get
\[
U\nT U^*=\oplus_{h=0}^{m-1}M_k(\mathbb{C}).
\]

We now propose another formulation with a slightly
different unitary operator in order to choose a non-trivial
Hamiltonian leading to a decoherence-free subalgebra
with non-trivial Hilbert space $\kk$ and non-trivial
multiplicity space $\mm$.
Let $(f_r)_{0\le r\le k-1}$ and $(g_h)_{0\le h\le m-1}$ be
orthonormal bases of $\mathbb{C}^k$ and $\mathbb{C}^m$.
For each $j$ let $j=mr +h$ the division of $j$ by $m$ with
remainder of $h$ and define the unitary
\[
F:\mathbb{C}^d \to \mathbb{C}^k\otimes \mathbb{C}^m,
\qquad
F v_j =  F v_{mr+h} = f_r\otimes g_h.
\]
It turns out that $FJ^kF^*=  \unit_{\mathbb{C}^k} \otimes S$
where $S$ is the unitary operator on $\mathbb{C}^m$ defined
by $Sg_{h}=\omega^{hk}g_{h}
=\hbox{\rm e}^{2\pi\mi h/m}g_{h}$,
and also $F J^n F^* =   \unit_{\mathbb{C}^k}\otimes S^{n/k}$.

The decoherence-free subalgebra $\nT$ is the commutant
of  $\{J^k,J^{-k}\}$ by Proposition \ref{prop:NT-circulant},
therefore it is isomorphic to the algebra operators $x\otimes y$
where $x\in M_k(\mathbb{C})$ and $y$ belongs to the commutant
of $S$. This is generated by operators
$x\otimes \left| g_h\right\rangle\left\langle g_h \right|$
$0\le h\le m-1$, therefore we recover the previous decomposition
\[
F\nT F^* \simeq \oplus_{i=0}^{ m-1} M_k(\mathbb{C})
\]
and $\kk_i=\mathbb{C}^k, \mm_i=\mathbb{C}$ for $i=0,\dots, m-1$.
Moreover, since $H=0$, we have $\mathcal{F}(\T)
=\nT=\{\,L_k,L_k^*\,\mid\, k=1,2\,\}^\prime$ and the
decoherence-free semigroup is the trivial semigroup of
identity maps.

However, if we consider the GKSL generator with
the above $L_1,L_2$ and Hamiltonian
\[
H = F^*\left( K\otimes \unit_{\mathbb{C}^m}
+ \unit_{\mathbb{C}^k}\otimes M_0\right)F
\]
with $K$ and $M_0$ self-adjoint on $\mathbb{C}^k$ and
$\mathbb{C}^m$ respectively and such that the commutant
of $\{\delta_{M_0}^{l}(S^{n/k}), \delta_{M_0}^{l}(S^{n/k})
\,\mid\, l\ge 0\,\}$ in $M_m(\mathbb{C})$ is trivial, then
\[
F\nT F^* \simeq M_k(\mathbb{C})\otimes \unit_{\mathbb{C}^m}
\]
is a factor and $\kk_1 = \mathbb{C}^k$, $\mm_1=\mathbb{C}^m$.

If $n/k$ and $m$ are coprime
(e.g. for $n=10$, $d=15$ so that $k=5,m=3$ and $n/k=2$),
then $S^{n/k}$ has non nondegenerate spectrum,
namely eigenvalues $\omega^{hn}=\hbox{\rm e}^{2\pi\mi h n/m k}
=\hbox{\rm e}^{2\pi\mi h n/d}$,  for $h=0,\ldots,m-1$,
are simple and we can write
\[
S^{n/k} = \sum_{h=0}^{m-1} \omega^{hn}\kb{g_h}{g_h}.
\]
Since the subalgebra of $M_m(\mathbb{C})$ generated by
$S^{n/k}$ is maximal abelian, any $X$ in the commutant
of $S^{n/k}$ which contains $\{\delta_{M_0}^{l}(S^{n/k}), \delta_{M_0}^{l}(S^{n/k}) \,\mid\, l\ge 0\,\}^\prime$
must be diagonal in the basis $(g_h)_{0\le h\le m}$.
We take, for instance,
\[
M_0=\sum_{h=0}^{m-1}
\left(\kb{g_{h+1}}{g_h}+\kb{g_{h-1}}{g_h}\right)
\]
so that $F ^*\unit_{\mathbb{C}^k}\otimes M_0 F v_{mr+h}
=v_{mr+h+1}+ v_{mr+h-1}$. Computing the commutator
$[M_0,S^{n/k}] $ we find
\begin{eqnarray*}
& = & \sum_{h,h'=0}^{m-1}
\left( \omega^{h'n}\kb{g_{h+1}}{g_h}\kb{g_{h'}}{g_{h'}}
+\omega^{h'n}\kb{g_{h-1}}{g_h}\kb{g_{h'}}{g_{h'}}\right) \\
& - & \sum_{h,h'=0}^{m-1}
\left( \omega^{h'n}\kb{g_{h'}}{g_{h'}}\kb{g_{h+1}}{g_h}
+\omega^{h'n}\kb{g_{h'}}{g_{h'}}\kb{g_{h-1}}{g_h}\right) \\
& = & \sum_{h=0}^{m-1}
\omega^{hn}\kb{g_{h+1}}{g_{h}}
+\sum_{h=0}^{m-1}\omega^{hn}\kb{g_{h-1}}{g_h}\\
& - & \sum_{h=0}^{m-1}
\omega^{(h+1)n}\kb{g_{h+1}}{g_h}
-\sum_{h=0}^{m-1}
\omega^{(h-1)n}\kb{g_{h-1}}{g_h} \\
& = &  \sum_{h=0}^{m-1}
\left( \omega^{hn}-\omega^{(h+1)n}\right)\kb{g_{h+1}}{g_h}
-\sum_{h=0}^{m-1}\left( \omega^{(h-1)n}-\omega^{hn}\right)
\kb{g_{h-1}}{g_h}
\end{eqnarray*}
An operator $X=\sum_h z_h \kb{g_h}{g_h}$
commutes with $[M_0,S^{n/k}]$ if and only if
\begin{eqnarray*}
[X,[M_0,S^{n/k}]]
& = & \sum_{h=0}^{m-1}
\left( \omega^{hn}-\omega^{(h+1)n}\right)
\left(z_{h+1}-z_h\right)\kb{g_{h+1}}{g_h} \\
& - & \sum_{h=0}^{m-1}\left( \omega^{(h-1)n}-\omega^{hn}\right)
\left(z_{h-1}-z_h\right)\kb{g_{h-1}}{g_h}= 0,
\end{eqnarray*}
By the linear independence of rank-one operators
$\kb{g_{h+1}}{g_h}, \kb{g_{h-1}}{g_h}$ ($h=0,\dots, d-1$),
$X$ commutes with $[M_0,S^{n/k}]$ if and only if
$z_h=z_{h+1}$ for all $h$, i.e. $X$ is a multiple of
the identity operator.

By Theorem \ref{th:main}, the decoherence-free
and decoherence-affected semigroups are now generated by
\[
\dfL = \mi [\,F^*(K\otimes \unit_{\mathbb{C}^m})F ,
\cdot \,] \qquad
\dL = \Ll - \dfL.
\]

\section*{Appendix}

First of all, we recall some preliminary definitions and results.
Given a von Neumann algebra $\M$, we denote by $\Pmin{\M}$ the set of its minimal projections. If $p$ is a projection in $\M$, its \emph{central support} $z_p$ is the smallest projection in the center $\Z{\M}$ of $\M$ such that $p\leq z_p$.
We refer to Takesaki (\cite{Take} Definition 5.9 p.155) for the following
definition.
\begin{definition}\label{def:atom}
Let $\M$ be a von Neumann algebra acting on $\h$.
$\M$ is called \emph{atomic} if for every non-zero projection $p\in\M$ there exists $q\in\Pmin{\M}$, $q\neq0$, such that $q\leq p$.
\end{definition}

Note that, since every projection $q\in p\M p$ is smaller than $p$, we have $p\in\Pmin{\M}$ if and only if $p\M p=\mathbb{C}p$.
\begin{lemma}\label{factor-atomic}
Let $\M$ be a type I factor. Then $\M$ is atomic.
\end{lemma}
\noindent{\bf Proof.}
By Theorem 4.2.1 in \cite{jones} we know that $\M$ is unitarily equivalent to $\mathcal{B}(\kk)\otimes\unit_\mm$ for some $\kk$ and $\mm$ Hilbert spaces. Since $\mathcal{B}(\kk)\otimes\unit_\mm$ is clearly an atomic algebra and every unitary isomorphism maps minimal projections into minimal projections, we can conclude that $\M$ is atomic too.
\hfill $\square$ \medskip

\begin{lemma}\label{minimal}
If $p$ is a non-zero minimal projection in $\M$, then its central support $z_p$ is a non-zero minimal projection in $\Z{\M}$.
\end{lemma}
\noindent{\bf Proof.}
By definition, $z_p$ is different from $0$. Let $q\in \Z{\M}$
be a non-zero projection such that $q\leq z_p$. Then $qp=pq=pqp
\leq pz_p p=p$, since $p$ and $q$ commute and $p\leq z_p$
by definition of central support. Now, the minimality of $p$ in
$\M$ implies either $qp=0$, i.e. $p\leq q^\perp$, or $qp=p$,
i.e. $p\leq q\leq z_p$. In this latter case we can conclude that
$q=z_p$ by definition of $z_p$. Otherwise, if $p\leq q^\perp$,
then $p=pz_p\leq z_pq^\perp\leq z_p$, so that
$z_p-q=z_pq^\perp=z_p$, since $z_pq^\perp$ is a projection
in $\Z{\M}$. As a consequence we have $q=0$,
which is a contradiction.
\hfill $\square$ \medskip

\begin{proposition}\label{atomic}
 Let $\M$ be a von Neumann algebra. The following are equivalent:
\begin{enumerate}
\item $\M$ is atomic;
\item there exists a collection $(p_i)_{i\in I}$ of mutually orthogonal projections in $\Pmin{\Z{\M}}$ such that $\sum_{i\in I}p_i=\unit$
and each $p_i\M p_i$ is a type I factor.
\end{enumerate}
\end{proposition}

\noindent{\bf Proof.}
$1.\Rightarrow 2.$\ Thanks to Lemma \ref{minimal} we can find a maximal family  $(p_i)_{i\in I}$ of mutually orthogonal minimal projections in $\Z{\M}$. Let $p=\sum_{i\in I}p_i$. If $p\neq \unit$, by the atomicity of $\M$ we can find a non-zero minimal projection $q\in\M$ such that $q\leq p^\perp$.
Denoting by $z_q$ the central support of $q$, by definition of $z_q$
we have $q\leq z_q\leq p^\perp$, since $p^\perp$ is a projection
in $\Z{\M}$ which majorizes $q$. Finally, $z_q$ is minimal in
$\Z{\M}$ by Lemma \ref{minimal}, contradicting the maximality
of  $(p_i)_{i\in I}$ and proving that $\sum_{i\in I}p_i=\unit$. \\
In order to  check that each $p_i\M p_i$ is a factor it is enough
to prove that its center,  which is a von Neumann algebra,
contains only trivial projections. So, let $q\in\Z{p_i\M p_i}$ be a
non-zero projection; since we have $0=[q,p_ix]=[p_iq,x]$ for all $x\in\M$, $p_iq$ belongs to $\Z{\M}$, and so $q=p_iq=p_iqp_i\in p_i\Z{\M}p_i=\mathbb{C}p_i$ by minimality of $p_i$ in $\Z{\M}$. We thus conclude that $q=p_i$, i.e.
$\Z{p_i\M p_i}=\mathbb{C}p_i$ and $p_i\M p_i$ is a factor.\\
Finally, since $\M$ is atomic, for every $i\in I$ there exists a non-zero minimal projection $q_i\in\M$ such that $q_i\leq p_i$;  therefore, each $q_i$ is a non-zero minimal projection in $p_i\M p_i$
and the factor $p_i\M p_i$ is type I.

$2.\Rightarrow 1.$\, First of all we note that
$\M=\oplus_{i\in I}p_i\M p_i$ since every $x\in\M$ can be written
as $x=\sum_{i\in I}p_ixp_i$,  because $p_i$ in $\Z{\M}$
and $\sum_ip_i=\unit$ .\\
We now show that $\M$ is atomic (Definition \ref{def:atom}).
Let $p\in\M$ be a non-zero projection, so that
$p=\sum_{i\in I}p_ipp_i$ with  $p_ipp_i$ a projection in
$p_i\M p_i$ for all $i\in I$. Since every $p_i\M p_i$
is a type I factor, it is atomic by Lemma \ref{factor-atomic};
hence, given $j\in I$ such that $p_jpp_j$ is not trivial,
we  can find a non-zero minimal projection $q\in p_j\M p_j$
with $q\leq p_jpp_j$. But $q$ is minimal in $\M$ too, and so
we can conclude that $q\leq p_jpp_j\leq\sum_{i\in I}p_ipp_i=p$.
\hfill $\square$ \medskip

\begin{remark}
Note that, if $\h$ is separable, then the index set $I$ introduced in the previous Lemma is necessarily countable, having
$\h=\oplus_{i\in I}p_i \h $.
\end{remark}
A necessary and sufficient condition  for $\M$ to be atomic
is given by the following result due to Tomiyama \cite{Tom3},
Theorem 5.
\begin{theorem}\label{Matomic}
Let $\M$ be a von Neumann algebra acting on $\h$. Then $\M$ is atomic if and only if there exists a normal conditional expectation $\E:\B\to\M$ such that $\Ran\E=\M$.
\end{theorem}

\section*{Acknowledgments}
The financial support of MIUR FIRB 2010 project RBFR10COAQ
{\it Quantum Markov Semigroups and their Empirical Estimation}
and of PRIN 2010–2011 project 2010MXMAJR–001
{\it Evolution differential problems: deterministic and stochastic approaches and their interactions} are gratefully acknowledged.

\bigskip
\noindent{\small JULIEN DESCHAMPS, Dipartimento di Matematica, Universit\`a di Genova, Via Dodecaneso 35, I-16146 Genova, Italy
\par\noindent{\tt deschamps@dima.unige.it}}
\medskip

\noindent{\small FRANCO FAGNOLA, Dipartimento di Matematica, Politecnico di Milano, Piazza Leonardo da Vinci 32, I-20133 Milano, Italy
\par\noindent{\tt franco.fagnola@polimi.it}}
\medskip

\noindent{\small EMANUELA SASSO, Dipartimento di Matematica, Universit\`a di Genova, Via Dodecaneso 35, I-16146 Genova, Italy
\par\noindent{\tt sasso@dima.unige.it}}
\medskip

\noindent{\small VERONICA UMANIT\`A, Dipartimento di Matematica, Universit\`a di Genova, Via Dodecaneso 35, I-16146 Genova, Italy
\par\noindent{\tt umanita@dima.unige.it}}

\end{document}